\documentclass[preprint,review,3p,times,onecolumn,authoryear]{elsarticle}

\usepackage{amssymb}
\usepackage{multicol}

\usepackage{lineno}

\journal{Computers \& Geosciences}

\usepackage{xspace,color,amsmath}
\usepackage{amssymb}
\usepackage{enumitem}
\usepackage{multirow}
\setitemize{noitemsep,topsep=0pt,parsep=0pt,partopsep=0pt}

\usepackage{etoolbox}
\newtoggle{finaldraft}
\togglefalse{finaldraft}

\begin{document}

\begin{frontmatter}



{
\title{Modeling electromagnetics on cylindrical meshes with applications to steel-cased wells}

\author[corresponding,UBC]{Lindsey J. Heagy}
\author[UBC]{Douglas W. Oldenburg}

\address[corresponding]{604-836-2715, lheagy@eos.ubc.ca}
\address[UBC]{Geophysical Inversion Facility, University of British Columbia}


\begin{abstract}
Simulating direct current resistivity, frequency domain electromagnetics and time domain electromagnetics in settings where steel cased boreholes are present is of interest across a range of applications including well-logging, monitoring subsurface injections such as hydraulic fracturing or carbon capture and storage. In some surveys, well-casings have been used as ``extended electrodes'' for near surface environmental or geotechnical applications. Wells are often cased with steel, which has both a high conductivity and a significant magnetic permeability. The large physical property contrasts as well as the large disparity in length-scales, which are introduced when a steel-cased well is in a modeling domain, makes performing an electromagnetic forward simulation challenging. Using this setting as motivation, we present a finite volume approach for modeling electromagnetic problems on cylindrically symmetric and 3D cylindrical meshes which include an azimuthal discretization. The associated software implementation includes modeling capabilities for direct current resistivity, time domain electromagnetics, and frequency domain electromagnetics for models that include variable electrical conductivity and magnetic permeability. Electric and magnetic fields, fluxes, and charges are readily accessible in any simulation so that they can be visualized and interrogated. We demonstrate the value of being able to explore the behaviour of electromagnetic fields and fluxes through examples which revisit a number of foundational papers on direct current resistivity and electromagnetics in steel-cased wells. The software implementation is open source and included as a part of the \texttt{SimPEG} software ecosystem for simulation and parameter estimation in geophysics.
\end{abstract}

}

\end{frontmatter}

\footnote{\textbf{Authorship statement}: Heagy developed the software and conducted the numerical experiments with guidance from Oldenburg who supervised the project. Heagy and Oldenburg contributed to the manuscript. }



\section{Introduction}

A number of geophysical electromagnetic (EM) problems lend themselves to cylindrical geometries. Airborne EM problems over a 1D layered earth or borehole-logging applications fall into this category; in these cases cylindrical modeling, which removes a degree of freedom in the azimuthal component, can be advantageous as it reduces the computation load. This is useful when running an inversion where many forward modelings are required, and it  is also valuable when exploring and building up an understanding of the behaviour of electromagnetic fields and fluxes in a variety of settings, such as the canonical model of an airborne EM sounding over a sphere, as it reduces feedback time between asking a question and visualizing results (e.g. \cite{Oldenburg2017}).

Beyond these simple settings, there are also a range of scenarios where the footprint of the survey is primarily cylindrical, but 2D or 3D variations in the physical property model may be present. For example if we consider a single sounding in an Airborne EM survey, the primary electric fields are rotational and the magnetic fields are poloidal, but the physical property model may have lateral variations or compact targets. More flexibility is required from the discretization to capture these features. In this case, a 3D cylindrical geometry, which incorporates azimuthal discretization may be advantageous. It allows finer discretization near the source where we have the most sensitivity and the fields are changing rapidly. Far from the source, the discretization is coarser, but it still conforms to the primary behaviour of the EM fields and fluxes and captures the rotational electric fields and poloidal magnetic flux.

In other cases, the most significant physical property variations may conform to a cylindrical geometry, for example in settings where vertical metallic well-casings are present, or in the emerging topic of using geophysics to ``look ahead'' of a tunnel boring machine. In particular, understanding the behavior of electromagnetic fields and fluxes in the presence of steel-cased wells is of interest across a range of applications, from characterizing lithologic units with well-logs \citep{Kaufman1990, Kaufman1993, Augustin1989}, to identifying marine hydrocarbon targets \citep{Kong2009, Swidinsky2013, Tietze2015}, to mapping changes in a reservoir induced by hydraulic fracturing or carbon capture and storage \citep{Pardo2013, Borner2015, Um2015, Weiss2016, hoversten2017borehole, Zhang2018}. Carbon steel, a material commonly used for borehole casings, is highly electrically conductive ($10^6 - 10^7$ S/m) and has a significant magnetic permeability ( $\geq 100$ $\mu_0$) \citep{wuhabashy1994}; it therefore can have a significant influence on electromagnetic signals. The large contrasts in physical properties between the casing and the geologic features of interest, along with the large range of scales that need to be considered to model both the millimeter-thick casing walls while also capturing geologic features, provide interesting challenges and context for electromagnetics in cylindrical geometries. As such, we will use EM simulations of conductive, permeable boreholes as motivation throughout this paper.

In much of the early literature, the casing was viewed as a nuisance which distorts the EM signals of interest. Distortion of surface direct current (DC) resistivity and induced polarization (IP) data, primarily in hydrocarbon settings, was examined in \citep{Wait1983, Holladay1984, Johnston1987} and later extended to grounded source EM and IP in \citep{Wait1985, Williams1985, Johnston1992}. Also in hydrocarbon applications, well-logging in the presence of steel cased boreholes is motivation for examining the behavior of electromagnetic fields and fluxes in the vicinity of casing. Initial work focussed on DC resistivity with \citep{Kaufman1990, Schenkel1990, Kaufman1993, Schenkel1994}, and inductive source frequency domain experiments with \citep{Augustin1989}. \cite{Kaufman1990} derives an analytical solution for the electric field at DC in an experiment where an electrode is positioned along the axis of an infinite-length well. The mathematical solutions presented shows how, and under what conditions, horizontal currents leak into the formation outside the well. Moreover, \cite{Kaufman1990} showed, based upon asymptotic analysis, which fields to measure inside the well so that information about the formation resistivity could be obtained. This analysis is extended to include finite-length wells in \cite{Kaufman1993}. \cite{Schenkel1994} show the importance of considering the length of the casing in borehole resistivity measurements, and demonstrate the feasibility of cross-well DC resistivity. They also show that the presence of a steel casing can improve sensitivity to a target adjacent to the well. In frequency domain EM, \cite{Augustin1989} consider a loop-loop experiment, where a large loop is positioned on the surface of the earth and a magnetic field receiver is within the borehole. Magnetic permeability is included in the analysis and a ``casing correction'', effectively a filter due to the casing on inductive-source data, is introduced. This work was built upon for considering cross-well frequency domain EM experiments \citep{Uchida1991, Wilt1996}.

For larger scale geophysical surveys, steel cased wells have been used as ``extended electrodes.'' \cite{Rocroi1985} used a pair of well casings as current electrodes for reservoir characterization in hydrocarbon applications. In near-surface settings \citep{Ramirez1996, Rucker2010, Rucker2012} considered the use of monitoring wells as current and potential electrodes for a DC experiment aimed at imaging nuclear waste beneath a leaking storage tank. Similarly, \cite{Ronczka2015} considers the use of groundwater wells for monitoring a saltwater intrusion and investigates numerical strategies for simulating casings as long electrodes. Imaging hydraulic fractures has been a motivator for a number of studies at DC or EM, among them \cite{Weiss2016, hoversten2017borehole}. Some of these have suggested the use of casings that include resistive gaps so that currents may be injected in a segment of the well and potentials measured across the other gaps along the well \citep{Nekut1995, Zhang2018}. There is also  interest in modeling casings for casing integrity applications where the aim of the DC or EM survey is to diagnose if a well is flawed or intact based on data collected on the surface \citep{Wilt2018}.

As computing resources increased, our ability to forward-simulate more complex scenarios has improved. However, the large physical property contrasts and disperate length scales introduced when a steel cased well is included in a model still present computational challenges. Even the DC problem, which is relatively computationally light, has posed challenges; those are exacerbated when solving the full Maxwell equations in the frequency (FDEM) or time domain (TDEM) and can become crippling for an inversion. For models where the source and borehole are axisymmetric, cylindrical symmetry may be exploited to reduce the dimensionality, and thus the number of unknowns, in the problem (e.g. \cite{Pardo2013, Heagy2015}).

To reduce computational load in a 3D simulation, a number of authors have employed simplifying assumptions. Several authors replaced the steel-cased well with a solid borehole, either with the same conductivity as the hollow-cased well (e.g. \cite{Um2015, Puzyrev2017}) or preserving the cross sectional conductance (e.g. \cite{Swidinsky2013, Kohnke2017}), so that a coarser discretization may be used; \cite{Haber2016} similarly replaces the borehole with a coarser conductivity structure and adopts an OcTree discretization locally refine the mesh around the casing. \cite{Yang2016} uses a circuit model and introduces circuit components to account for the steel cased well in a 3D DC resistivity experiment; \cite{Weiss2017} adopts a similar strategy in a Finite Element scheme. Another approach has been to replace the well with an ``equivalent source'', for example, a collection of representative dipoles, inspired from \cite{cuevas2014}, or with a line-charge distribution for a DC problem \citep{Weiss2016}. For the frequency domain electromagnetic problem, a method of moments approach, which replaces the casing with a series of current dipoles, has been taken in \cite{Kohnke2017}.

For 3D survey geometries, only a handful of forward simulations which accurately discretize the casing have been demonstrated, and they have been achieved at significant computational cost. Recent examples, including \cite{Commer2015, Um2015, Puzyrev2017}, perform time and frequency domain simulations with finely-discretized boreholes; they required the equivalent of days of compute-time for a single forward simulation to complete. While these codes will undoubtedly see improvements in efficiency, what we present here is an alternative approach to the discretization which capitalizes on the cylindrical geometry of a borehole. Thus far, the majority of the literature has focussed on electrical conductivity, with little attention being paid to magnetic permeability, leaving open many questions about how it alters the behavior of the fields and fluxes and the impact it has on data measured at the surface. In our approach, we ensure that variable magnetic permeability can be included in order to facilitate exploration of these questions.

We introduce an approach and associated open-source software implementation for simulating Maxwell's equations over conductive, permeable models on 2D and 3D cylindrical meshes. The software is written in Python \citep{van1995python} and is included as an extension to the \texttt{SimPEG} ecosystem \citep{Cockett2015, Heagy2017}. Within the context of current research connected to steel-cased wells, our aim with the development and distribution of this software is two-fold: (1) to facilitate the exploration of the physics of EM in these large-contrast settings, and (2) to provide a simulation tool that can be used for testing other EM codes. The large physical property contrasts in both conductivity and permeability means the physics is complicated and often non-intuitive; as such, we ensure that the researcher can readily access and visualize fields, fluxes, and charges in the simulation domain. This is particularly useful when the software is used in conjunction with Jupyter notebooks which facilitate exploration of numerical results \citep{Perez2015}. As the mesh conforms to the geometry of a vertical borehole, a fine discretization can be used in its vicinity without resulting in a onerous computation. This provides the opportunity to build an understanding of the physics of EM in settings with vertical boreholes prior to moving to settings with deviated and horizontal wells. We demonstrate the software with examples at DC, in the frequency domain, and in the time domain. Source-code for all examples is provided as Jupyter notebooks at https://github.com/simpeg-research/heagy-2018-emcyl \citep{Heagy2018}; they are licensed under the permissive MIT license with the hope of reducing the effort necessary by a researcher to compare to or build upon this work.

Our paper is organized in the following manner. In section \ref{sec:numerical_tools}, we introduce the governing equations, Maxwell's equations, and describe their discretization in cylindrical coordinates. We then compare our numerical implementation to the finite element and finite difference results shown in \citep{Commer2015} as well as a finite volume OcTree simulation described in \citep{Haber2007}. Section \ref{sec:numerical_examples} contains numerical examples of the DC, frequency domain EM, and time domain EM implementations. The two DC resistivity examples (sections \ref{sec:dc_resistivity_part1} and \ref{sec:dc_resistivity_part2}) are built upon the foundational work in \citep{Kaufman1990, Kaufman1993} which use asymptotic analysis to draw conclusions about the behavior of the electric fields, currents, and charges for a well where an electrode has been positioned along its axis. The next example, in section \ref{sec:TDEM}, is motivated by the interest in using a steel-cased well as an ``extended electrode'' in a time domain EM experiment. We perform a ``top-casing'' experiment, with one electrode connected to the top of the well and examine the currents in the surrounding geologic formation through time. Our final two examples, in sections \ref{sec:FDEM_part1} and \ref{sec:FDEM_part2}, consider a frequency domain experiment inspired by \citep{Augustin1989}. These examples demonstrate the impact of magnetic permeability on the character of the magnetic flux within the vicinity of the borehole and discusses the resulting magnetic field measurements made within a borehole.
\section{Methodology}
\label{sec:numerical_tools}
\subsection{Governing Equations}
The governing equations under consideration are Maxwell's equations. Here we provide a brief overview and recommend \cite{Ward1988} for more detail. Under the quasi-static approximation, Maxwell's equations are given by:
\begin{equation}
\begin{split}
\nabla \times \vec{e} + \frac{\partial \vec{b}}{\partial t} = 0 \\
\nabla \times \vec{h} - \vec{j} = \vec{s}_e
\end{split}
\label{eq:MaxwellTime}
\end{equation}
where $\vec{e}$ is the electric field, $\vec{b}$ is the magnetic flux density, $\vec{h}$ is the magnetic field, $\vec{j}$ is the current density and $\vec{s_e}$ is the source current density. Maxwell's equations can also be formulated in the frequency domain, using the $e^{i \omega t}$ Fourier Transform convention, they are
\begin{equation}
\begin{split}
\nabla \times \vec{E} + i\omega\vec{B} &= 0 \\
\nabla \times \vec{H} - \vec{J} &= \vec{S}_e
\end{split}
\label{eq:MaxwellFreq}
\end{equation}
The fields and fluxes are related through the physical properties: electrical conductivity ($\sigma$, or its inverse, resistivity $\rho$) and magnetic permeability ($\mu$), as described by the constitutive relations
\begin{equation}
\begin{split}
\vec{J} = \sigma \vec{E} \\
\vec{B} = \mu \vec{H}
\end{split}
\label{eq:ConstitutiveRelations}
\end{equation}
At the zero-frequency limit, we also consider the DC resistivity experiment, described by
\begin{equation}
\begin{split}
\nabla \cdot \vec{j} &= I\left(\delta(\vec{r} - \vec{r}_{s^{+}}) - \delta(\vec{r} - \vec{r}_{s^{-}})\right) \\
\vec{e} &= - \nabla \phi
\end{split}
\label{eq:DCequations}
\end{equation}
where $I$ is the magnitude of the source current density, $\vec{r}_{s^+}$ and $\vec{r}_{s^-}$ are the locations of the current electrodes, and $\phi$ is the scalar electric potential.

Of our numerical tools, we require the ability to simulate large electrical conductivity contrasts, include magnetic permeability, and solve Maxwell's equations at DC, in frequency and in time in a computationally tractable manner. Finite volume methods are advantageous for modeling large physical property contrasts as they are conservative and the operators ``mimic'' properties of the continuous operators, that is, the edge curl operator is in the null space of the face divergence operator, and the nodal gradient operator is in the null space of the edge curl operator \citep{Hyman1999}. As such, they are common practice for many electromagnetic simulations (e.g. \cite{Horesh2011, Haber2014, Jahandari2014} and references within), and will be our method of choice.
\subsection{Finite Volume Discretization}
\begin{figure}
    \begin{center}
    \includegraphics[width=\columnwidth]{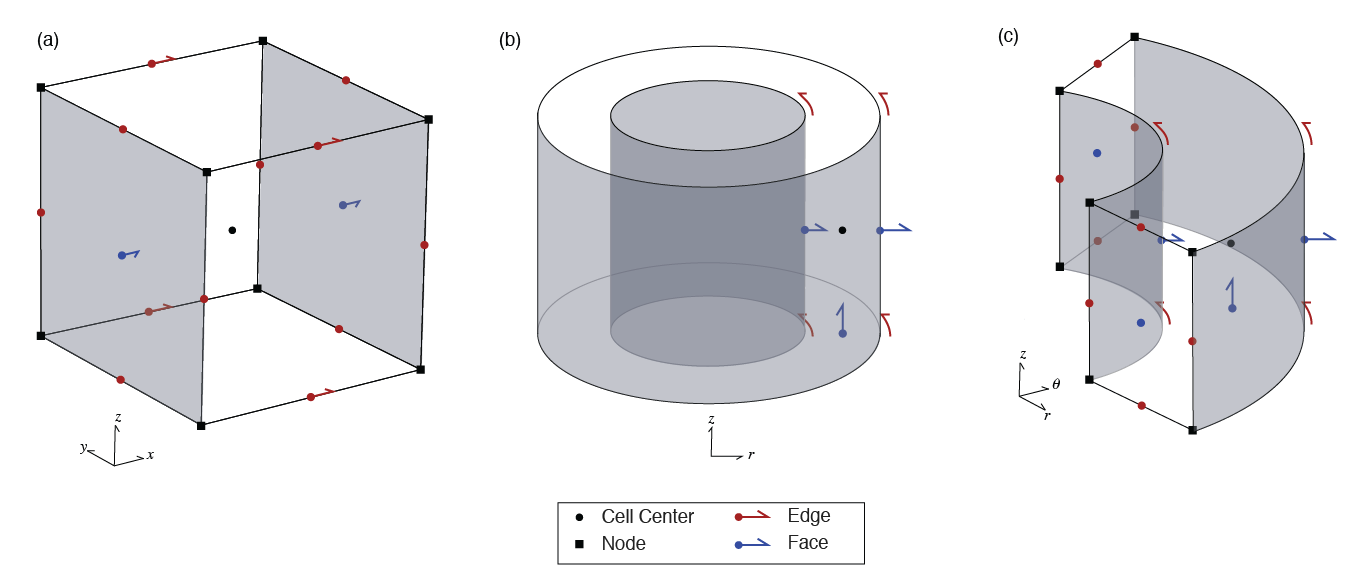}
    \end{center}
\caption{
    Anatomy of a finite volume cell in a (a) cartesian,
    regtangular mesh, (b) cylindrically symmetric mesh, and
    (c) a three dimensional cylindrical mesh.
}
\label{fig:CylFiniteVolume}
\end{figure}

To represent a set of partial differential equations on the mesh, we use a staggered-grid approach \citep{Yee1966} and discretize fields on edges, fluxes on faces, and physical properties at cell centers, as shown in Figure \ref{fig:CylFiniteVolume}. Scalar potentials can be discretized at cell centers or nodes. We consider both cylindrically symmetric meshes and fully 3D cylindrical meshes; the anatomy of a finite volume cell for these scenarios is shown in Figure \ref{fig:CylFiniteVolume} (b) and (c).

To discretize Maxwell's equations in the time domain (equation \ref{eq:MaxwellTime}) or in the frequency domain (equation \ref{eq:MaxwellFreq}), we invoke the constitutive relations to formulate our system in terms of a single field and a single flux. This gives a system in either the electric field and magnetic flux (E-B formulation), or the magnetic field and the current density (H-J formulation). For example, in the frequency domain, the E-B formulation is
\begin{equation}
    \begin{split}
        \mathbf{C} \mathbf{e} + i\omega\mathbf{b} &= \mathbf{0} \\
        \mathbf{C}^\top \mathbf{M}_{\boldsymbol{\mu}^{-1}}^f \mathbf{b} - \mathbf{M}_{\boldsymbol{\sigma}}^e \mathbf{e} &= \mathbf{s_e}
    \end{split}
    \label{eq:DiscreteFDEMEB}
\end{equation}
and the H-J formulation is
\begin{equation}
    \begin{split}
        \mathbf{C}^\top \mathbf{M}_{\boldsymbol{\rho}}^f \mathbf{j} + i\omega\mathbf{M}_{\boldsymbol{\mu}}^e\mathbf{h} &= \mathbf{0} \\
        \mathbf{C} \mathbf{h} - \mathbf{j} &= \mathbf{s_e}
    \end{split}
    \label{eq:DiscreteFDEMHJ}
\end{equation}
where $\mathbf{e}, \mathbf{b}, \mathbf{h}, \mathbf{j}$ are vectors of the discrete EM fields and fluxes; $\mathbf{s_m}$ and $\mathbf{s_e}$ are the discrete magnetic and electric source terms, respectively; $\mathbf{C}$ is the edge curl operator, and the matrices $\mathbf{M}_{\text{prop}}^{e,f}$ are the edge / face inner product matrices. In particular, variable electrical conductivity and variable magnetic permeability are captured in the discretization. The time domain equations are discretized in the same manner as is discussed in \citep{Heagy2017}; for time-stepping, a first-order backward Euler approach is used. Although the midpoint method, which is second-order accurate, could be considered, it is susceptible to oscillations in the solution, which reduce the order of accuracy, unless a sufficiently small time-step is used \citep{Haber2004, Haber2014}.

At the zero-frequency limit, each formulation has a complementary discretization for the DC equations, for the E-B formulation the discretization leads to a nodal discretization of the electric potential $\boldsymbol{\phi}$, giving
\begin{equation}
    \begin{split}
        - \mathbf{G}^\top \mathbf{M}_{\boldsymbol{\sigma}}^e \mathbf{e} &= \mathbf{q} \\
        \mathbf{e} &= -\mathbf{G}\boldsymbol{\phi}
    \end{split}
    \label{eq:DiscreteDCNodal}
\end{equation}
where $\mathbf{G}$ is the nodal gradient operator, and $\mathbf{q}$ is the source term, defined on nodes. Note that the nodal gradient takes the discrete derivative of nodal variables, and thus the output is on edges. The H-J formulation leads naturally to a cell centered discretization of the electric potential
\begin{equation}
    \begin{split}
        \mathbf{V} \mathbf{D}  \mathbf{j} &= \mathbf{q} \\
        \mathbf{M}_{\boldsymbol{\rho}}^f \mathbf{j} &= \mathbf{D}^\top \mathbf{V} \boldsymbol{\phi}
    \end{split}
    \label{eq:DiscreteDCCC}
\end{equation}
Where $\mathbf{D}$ is the face divergence operator, $\mathbf{V}$ is a diagonal matrix of the cell volumes, $\mathbf{q}$ is the source term, which is  defined at cell centers as is $\boldsymbol{\phi}$. Here, the face divergence takes the discrete derivative from faces to cell centers, thus its transpose takes a variable from cell centers to faces. For a tutorial on the finite volume discretization of the DC equations, see \citep{Cockett2016}.

For the EM simulations, natural boundary conditions are employed; in the E-B formulation, this means $\vec{B}\times\vec{n} = 0\vert_{\partial \Omega}$, and in the H-J formulation, we use $\vec{J}\times\vec{n} = 0\vert_{\partial \Omega}$. Within the DC simulations, there is flexibility on the choice of boundary conditions employed. In the simplest scenario, for the nodal discretization, we use Neumann boundary conditions, $\sigma\vec{E} \cdot \vec{n} = 0\vert_{\partial \Omega}$, and for the cell centered discretization, we use Dirichlet boundary conditions $\phi = 0\vert_{\partial \Omega}$.

When employing a cylindrical mesh, the distinction between where the electric and magnetic contributions are discretized in each formulation has important implications. If we consider the cylindrically symmetric mesh (Figure \ref{fig:CylFiniteVolume}b) and a magnetic dipole source positioned along the axis of symmetry (sometimes referred to as the TE mode), we must use the E-B formulation of Maxwell's equation to simulate the resulting toroidal magnetic flux and rotational electric fields. If instead, a vertical current dipole is positioned along the axis of symmetry (also referred to as the TM mode), then the H-J formulation of Maxwell's equations must be used in order to simulate toroidal currents and rotational magnetic fields. The advantage of a fully 3D cylindrical mesh provides additional degrees of freedom, with the discretization in the azimuthal direction, allowing us to simulate more complex responses. However, in order to avoid the need for very fine discretization in the azimuthal direction, we should select the most natural formulation of Maxwell's equations given the source geometry being considered. For a vertical steel cased well and a grounded source, we expect the majority of the currents to flow vertically and radially, thus the more natural discretization to employ is the H-J formulation of Maxwell's equations.

\begin{figure}
    \begin{center}
    \includegraphics[width=0.9\columnwidth]{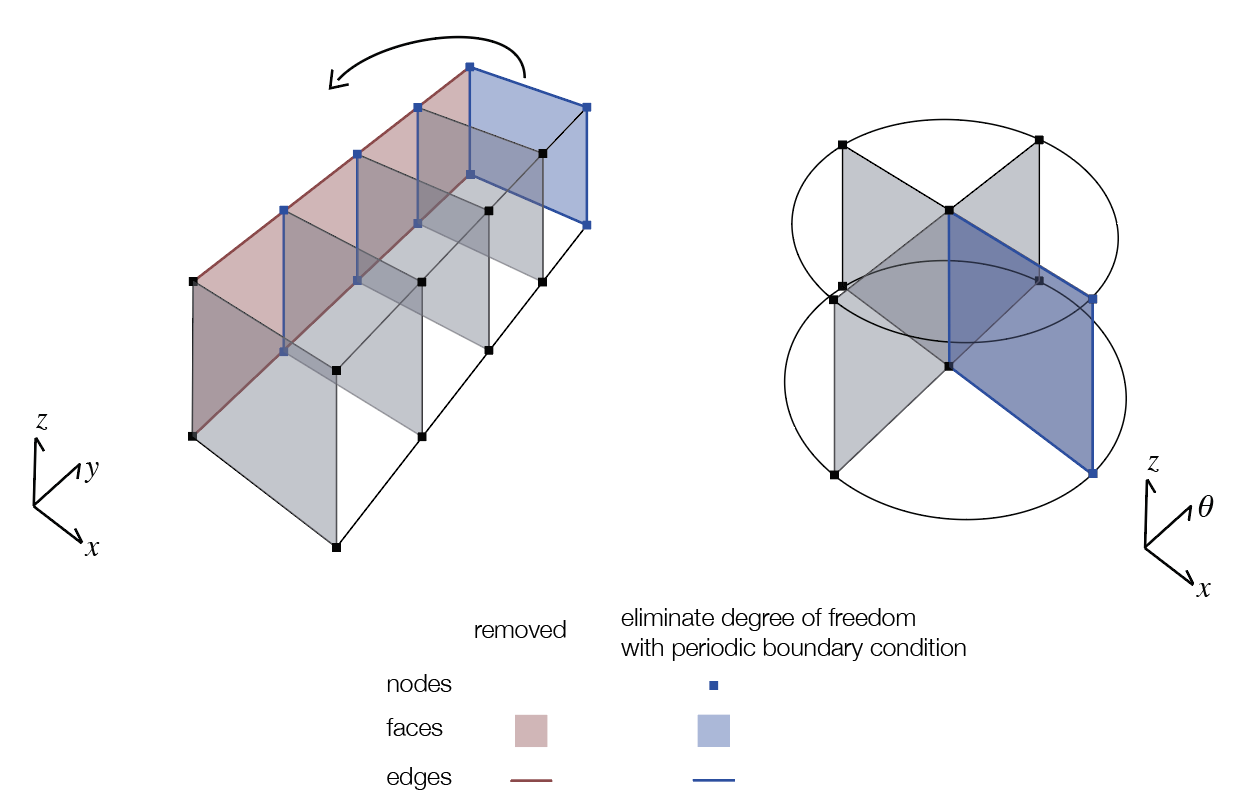}
    \end{center}
\caption{Construction of a 3D cylindrical mesh from a cartesian mesh.}
\label{fig:cylwrap}
\end{figure}

\cite{Haber2014} provides derivations and discussion of the differential operators and inner product matrices; though they are described for a cartesian coordinate system and a rectangular grid, the extension to a three dimensional cylindrical mesh is straightforward. Effectively, a cartesian mesh is wrapped so that the $x$ components become $r$ components, and $y$ components become $\theta$ components, as shown in Figure \ref{fig:cylwrap}.

The additional complications that are introduced are: (1) the periodic boundary condition introduced on boundary faces and edges in the azimuthal direction, (2) the removal of radial faces and azimuthal edges along the axis of symmetry, and (3) the elimination of the degrees of freedom of the nodes and edges at the boundary and as well as the nodes and vertical edges along the axis of symmetry. The implementation of the 3D cylindrical mesh is provided as a part of the \texttt{discretize} package (http://discretize.simpeg.xyz), which is an open-source python package that contains finite volume operators and utilities for a variety of mesh-types. All differential operators are tested for second order convergence and for preservation of mimetic properties (as described in \cite{Haber2014}). \texttt{discretize} is developed in a modular, object-oriented manner and interfaces to all of the \texttt{SimPEG} forward modeling and inversion routines, thus, once the differential operators have been implemented, they can be readily used to perform forward simulations \citep{Cockett2015, Heagy2017}.  One of the benefits of \texttt{SimPEG} for forward simulations is that values of the fields and fluxes are readily computed and visualized, which enables researchers to examine the physics as well as to simulate data. Development within the \texttt{SimPEG} ecosystem follows best practices for modern, opens-source software, including: peer review of code changes and additions, versioning, automated testing, documentation, and issue tracking.

\subsection{Validation}
Testing for the DC, TDEM, and FDEM implementations includes comparison with analytic solutions for a dipole in a whole-space. These examples are included as supplementary examples with the distributed notebooks. We have also compared the cylindrically symmetric implementation at low frequency with a DC simulations from a Resistor Network solution developed in MATLAB with (Figure 3 in \cite{Yang2016}).

Here, we include a comparison with the time domain electromagnetic simulation shown in Figures 13 and 14 of \cite{Commer2015}. A 200m long well, with a conductivity of $10^{6}$ S/m, outer diameter of 135 mm, and casing thickness of 12 mm is embedded in a 0.0333 S/m background. For the material inside the casing, we use a conductivity equal to that of the background. The conductivity of the air is set to $3 \times 10^{-4}$ S/m and the permeability of the casing is ignored ($\mu = \mu_0$). A 10 m long inline electric dipole source is positioned on the surface, 50 m radially from the well. The radial electric field is sampled at 5 m, 10 m, 100 m, 200 m and 300 m along a line $180^{\circ}$ from the source.

Two simulations are included in \cite{Commer2015}: a finite element (FE) and a finite difference (FD) solution. Both simulation meshes capture the thickness of the casing with a single cell or single tetrahedral element. Additionally, we include a comparison with the 3D UBC finite volume OcTree time domain code \citep{Haber2007}. The OcTree mesh allows for adaptive refinement of the mesh around sources, receivers, and conductivity structures within the domain, thus reducing the number of unknowns in the domain as compared to a tensor mesh.

\begin{figure}
    \begin{center}
    \includegraphics[width=\columnwidth]{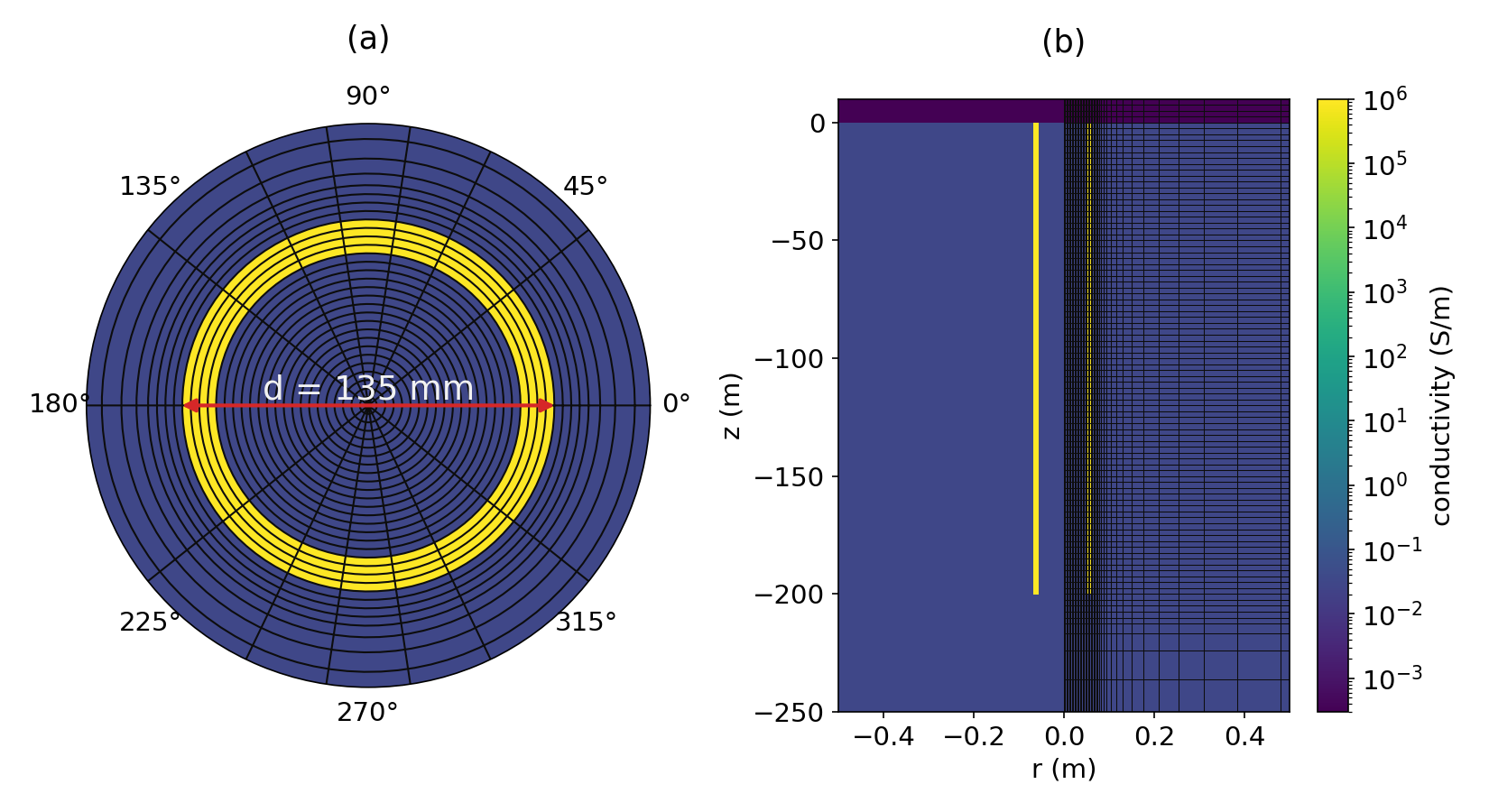}
    \end{center}
\caption{
    Depth slice (left) and cross section (right) through the 3D cylindrical
    mesh used for the comparison with \cite{Commer2015}.
    The source and recievers are positioned along the $\theta = 90^\circ$ line.
    The mesh extends 17km raidally and 30km vertically to ensure that the fields
    have sufficiently decayed before reaching the boundaries.
}
\label{fig:commer_model}
\end{figure}

For the 3D cylindrical simulation (SimPEG), we use a mesh that has 4 cells radially across the width of the casing, 2.5m vertical discretization, and azimuthal refinement near the source and receivers (along the $\theta=90^\circ$ line), as shown in Figure \ref{fig:commer_model}. To solve the system-matrix, the direct solver PARDISO was used \citep{Petra2014, Cosmin2016}. The simulation took 14 minutes to run on a single Intel Xeon X5660 processor (2.80GHz). The details of each simulation are shown in Table \ref{tab:commer_comparison}.

\begin{table}
\centering
    \small
    \begin{tabular}[htb]{| p{2cm} | p{2.5cm} | p{3cm} | p{3.75cm} | p{2.25cm} |}
        \hline
         & \textbf{Mesh} & \textbf{Timestepping} & \textbf{Compute Resources} & \textbf{Compute Time} \\
        \hline
        Commer FE & 8 421 559 tetrahedral elements & 893 time steps \newline 9 factorizations & single core \newline  Intel Xeon X5550 (2.67 GHz) & 63 hours\\
        \hline
        Commer FD & 2 182 528 cells & $\Delta t = 3 \times 10^{-10}$ s \newline 120 598 277 time-steps & 512 cores \newline Intel Xeon (2.33 GHz) & 23.2 hours\\
        \hline
        UBC OcTree & 5 011 924 cell & 154 time steps \newline  10 factorizations &  single core \newline Intel Xeon X5660 (2.80 GHz) & 57 minutes\\
        \hline
        SimPEG & 314 272 cells & 187 time-steps  \newline 7 factorizations & single core \newline Intel Xeon X5660 (2.80GHz) & 14 minutes \\
        \hline
    \end{tabular}
    \caption{
        Simulation details for the results shown in Figure \ref{fig:commer_results}.
        Note that the discretizations in the Commer FE and FD codes use one element
        or one cell across the width of the casing, as does the UBC code.
        The SimPEG simulation uses 4 cells across the width of the casing.
        For the time-stepping, each chance in step length requires a matrix factorization.
        Values for the Commer FE and FD solutions are from \cite{Commer2015, Um2015}.
    }
    \label{tab:commer_comparison}
 \end{table}

\begin{figure}[htb]
    \begin{center}
    \includegraphics[width=0.8\columnwidth]{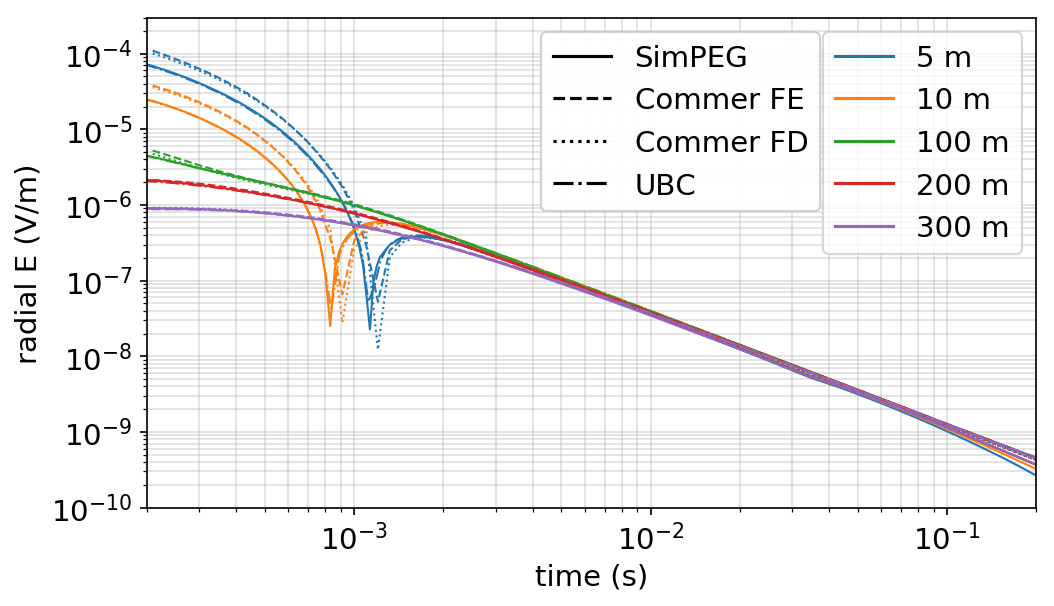}
    \end{center}
\caption{Time domain EM response comparison with \citep{Commer2015}. Each of the different line colors is associated with a different location; offsets are with respect to the location of the well.}
\label{fig:commer_results}
\end{figure}

In Figure \ref{fig:commer_results}, we show the absolute value of the radial electric field sampled at fives stations; each of the different line colors is associated with a different location, and offsets are with respect to the location of the well. Solutions were interpolated to the same offset using nearest neighbor interpolation.The 3D cylindrical simulation (SimPEG) is plotted with a solid line and overlaps with the UBC solution (dash-dot line) for all times shown. The finite element (FE) solution from \cite{Commer2015} is shown with the dashed lines, and the finite difference (FD) solution is plotted with dotted lines. The 3D cylindrical (SimPEG) and UBC solutions are overall in good agreement with the solutions from \cite{Commer2015}. There is a difference in amplitude and position of the zero-crossing (the v-shape visible in the blue and orange curves) between the Commer solutions and the SimPEG / UBC solutions at the shortest two offsets in the early times. At such short offsets from a highly conductive target, details of the simulation and discretization, such as the construction of the physical property matrices in each of the various approaches become significant; this likely accounts for the discrepancies but a detailed code-comparison is beyond the scope of this paper. Our aim with this comparison is to provide evidence that our numerical simulation is performing as expected; the overall agreement with Commer's and UBC's results is provides confidence that it is.

This example demonstrates agreement between the 3D cylindrical solution and solutions obtained with independently developed codes. Importantly, it also shows how, by using a cylindrical discretization which conforms to the conductivity structure of interest, the size of the mesh and resultant cost of the computation can be greatly reduced. This is true even with relatively conservative spatial and temporal discretizations. Minimizing computation time was not a main focus in the development of the software and there are still opportunities for improving efficiency. As an open-source project, contributions from the wider community are encouraged.
\section{Numerical Examples}
\label{sec:numerical_examples}

We demonstrate the implementation through examples using the DC, time domain EM and frequency domain EM codes. To focus discussion, each of the examples explores an aspect of the physical behavior of electromagnetic fields and fluxes in the presence of a steel-cased well.
\subsection{DC Resistivity Part 1: Electric fields, currents and charges in a long well}
\label{sec:dc_resistivity_part1}

In his two seminal papers on the topic, Kaufman uses transmission line theory to draw conclusions about the behaviour of the electric field when an electrode is positioned inside of an infinite casing. In this first example, we will revisit some of the physical insights discussed in \citep{Kaufman1990, Kaufman1993} that followed from an analytical derivation and compare those to our numerical results. In the second example, we look at the distribution of current and charges as the length of the well is varied and compare those to the analytical results discussed in \citep{Kaufman1993}

We start by considering a 1km long well ($10^6$ S/m) in a whole space ($10^{-2}$ S/m), with the conductivity of the material inside the borehole equal to that of the whole space.  For modeling, we will use a cylindrically symmetric mesh. The positive electrode is positioned on the borehole axis in the mid-point of a 1km long well;  a distant return electrode is positioned 1km away at the same depth.

\begin{figure}
    \begin{center}
    \includegraphics[width=0.7\columnwidth]{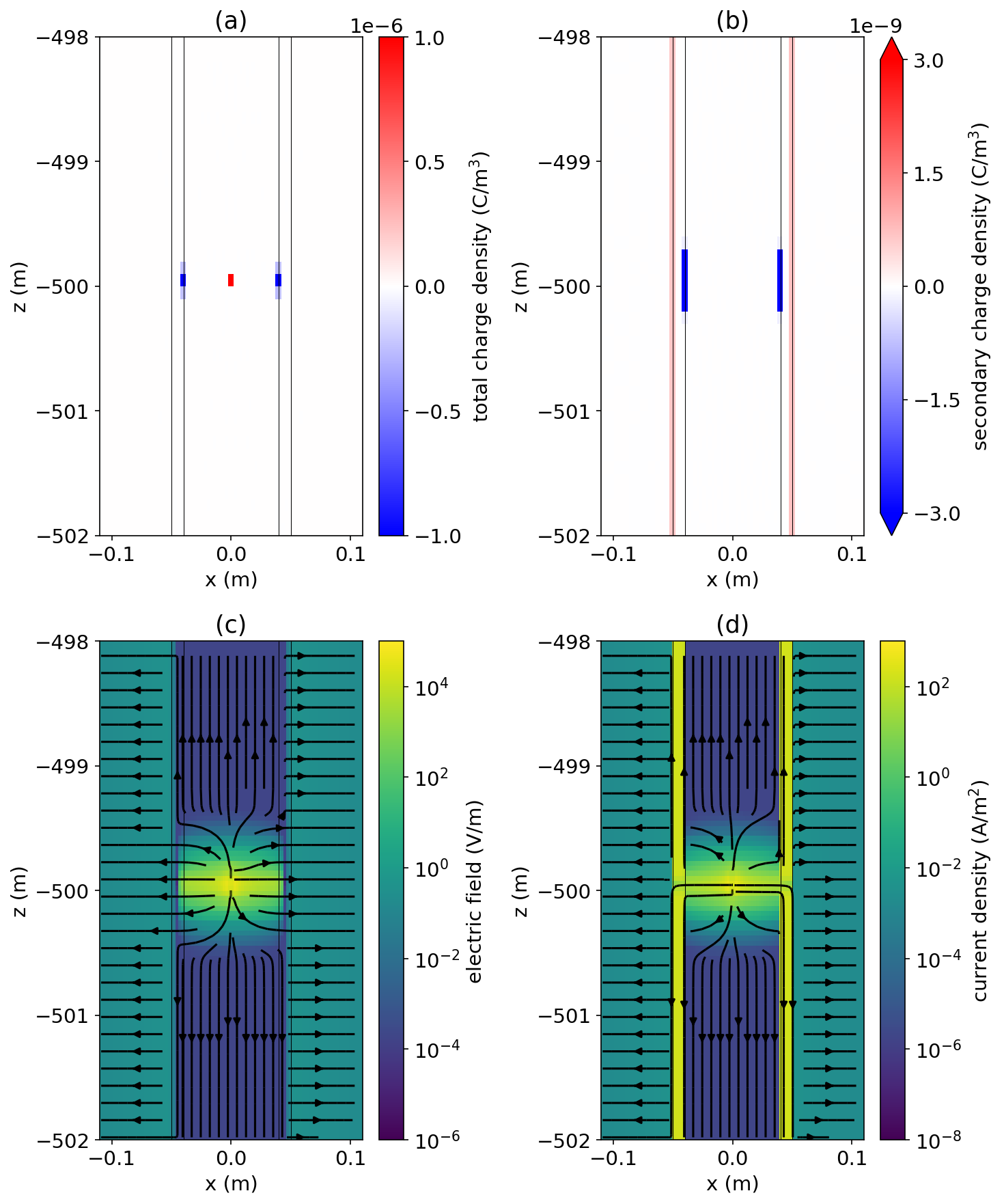}
    \end{center}
\caption{(a) Total charge density, (b) secondary charge density, (c) electric field, and (d) current density in a section of the pipe near the source at z=-500m.}
\label{fig:kaufman_zones}
\end{figure}

Kaufman discusses the behavior of the electric field by dividing the response into three zones: a near zone, an intermediate zone and a far zone \citep{Kaufman1990, Kaufman1993}. In the near zone, the electric field has both radial and vertical components, negative charges are present on the inside of the casing, and positive charges are present on the outside of the casing. The near zone is quite localized and typically, its vertical extent is no more than $\sim 10$ borehole radii away from the electrode. To examine these features in our numerical simulation, we have plotted in Figure \ref{fig:kaufman_zones}: (a)  the total charge, (b) secondary charges, (c) electric field, and (d) current density in a portion of the model near the source. The behaviours expected by Kaufman are consistent with our numerical results.

Within the near-zone, the total charge is dominated by the large positive charge at the current electrode location and negative charges that exist along the casing wall where current is moving from a resistive region inside the borehole into a conductor. The extent of the negative charges along the inner casing wall is more evident when we look at the secondary charge, which is obtained by subtracting the charge that would be observed in a uniform half-space from the total charge (Figure \ref{fig:kaufman_zones}b). Inside the casing, we can see the transition from near-zone behavior to intermediate zone behavior approximately 0.5 m above and below the source; that is equal to 10 borehole radii from the source location, which agrees with Kaufman's conclusion.

In the intermediate zone, Kaufman discusses a number of interesting aspects with respect to  the behavior of the electric fields and currents which we can compare with the observed behavior in Figure \ref{fig:kaufman_zones}. Among them, he shows that the electric field within the borehole and casing is directed along the vertical axis; as a result no charges accumulate on the inner casing wall. Charges do, however, accumulate on the outer surface of the casing; these  generate radially-directed electric fields and currents, often referred to as “leakage currents”, within the formation. At each depth slice through the casing and borehole, the electric field is uniform, however, due to the high conductivity of the casing, most of the current flows within the casing.  The vertical extent of the intermediate zone depends on the resistivity contrast between the casing and the surrounding formation and extends beyond several hundred meters before transitioning to the far zone, where the influence of the casing disappears \citep{Kaufman1990}.

The radially directed fields from the casing, and the length of the intermediate zone, have practical implications in the context of well-logging because they delineate the region in which measurements can be made to acquire information about the formation resistivity outside the well. Within the intermediate zone, fields behave like those due to a transmission line \citep{Kaufman1990}, and multiple authors have adopted modeling strategies that approximate the well and surrounding medium as a transmission line \citep{Kong2009, Aldridge2015}. We will extend this analysis in the next example and discuss how the length of the well impacts the behavior of the charges, fields, and fluxes.
\subsection{DC Resistivity Part 2: Finite Length Wells}
\label{sec:dc_resistivity_part2}

In \citep{Kaufman1993}, the transmission-line analysis was extended to consider finite-length wells. Inspired by the interest in using the casing as an ``extended electrode'' for delivering current to depth (e.g. \cite{Schenkel1994, Um2015, Weiss2016, hoversten2017borehole}), here we consider a 3D DC resistivity experiment where one electrode is connected to the top of the well. We will examine the current and charge distribution for wells ranging in length from 250 m to 4000 m and compare those to the observations in \citep{Kaufman1993}. The conductivity of the well is selected to be $10^6$ S/m. A uniform background conductivity of $10^{-2}$ S/m is used and the return electrode is positioned 8000m from the well; this is sufficiently far from the well that we do not need to examine the impact of the return electrode location in this example. A 3D cylindrical mesh was used for the simulation.

\begin{figure}[htb]
    \begin{center}
    \includegraphics[width=\columnwidth]{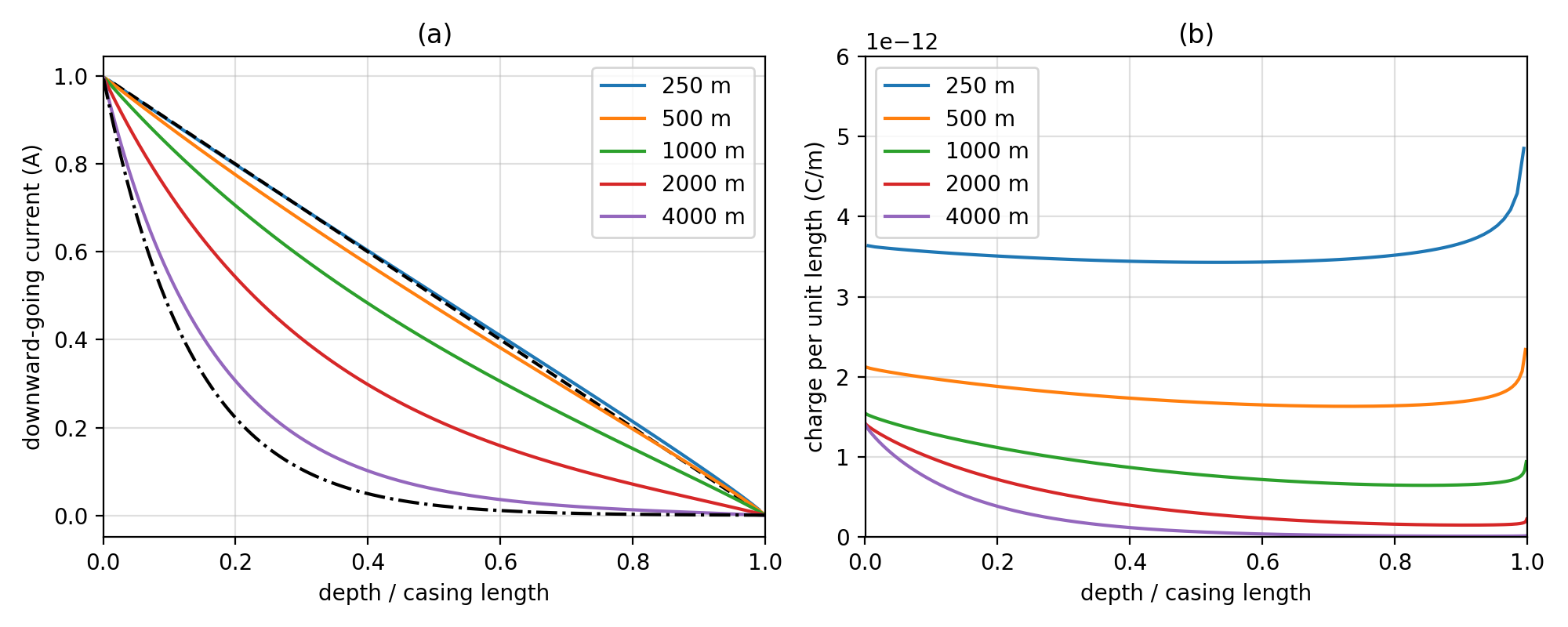}
    \end{center}
\caption{
    (a) Current along a well for 5 different wellbore lengths.
    The x-axis is depth normalized by the length of the well. The black
    dashed line shows the short-well approximation (equation 45 in \cite{Kaufman1993})
    for a 200m long well. The black dash-dot line shows the long-well approximation
    (equation 53 in \cite{Kaufman1993}) for a 4000m well.
    (b) Charge per unit length along the well for 5 different wellbore lengths.
}
\label{fig:kaufman_finite_well}
\end{figure}

\cite{Kaufman1993} derives a solution for the current within a finite length well and discusses two end-member cases: a short well and a long well. ``Short'' versus ``long'' are defined on the product of $\alpha L_c$, where $L_c$ is the length of the casing and $\alpha = 1/\sqrt{S T}$, where $S$ is the cross-sectional conductance of the casing and has units of S$\cdot$m ($S = \sigma_c 2\pi a \Delta a$, for a casing with radius $a$ and thickness $\Delta a$), and $T$ is the transverse resistance. The transverse resistance  is approximately equal to the resistivity of the surrounding formation (for more discussion on where this approximation breaks down, see \cite{Schenkel1994}). For short wells, $\alpha L_c \ll 1$, the current decreases linearly with distance, whereas for long wells, where $\alpha L_c \gg 1$, the current decays exponentially with distance from the source, with the rate of decay being controlled by the parameter $\alpha$. In Figure \ref{fig:kaufman_finite_well} (a), we show current in the well for 5 different borehole lengths. The x-axis is the distance from the source normalized by the length of the well. We also show the two end-member solutions (equations 45 and 53) from \cite{Kaufman1993}. There is significant overlap between the 250m numerical solution and the short well approximation. As the length of the well increases, exponential decay of the currents becomes evident. Since $\alpha$ is quite small, for this example $\alpha = 2 \times 10^{-3} m^{-1}$, the borehole must be very long to reach the other end member which corresponds to the exponentially decaying solution.

In Figure \ref{fig:kaufman_finite_well} (b), we have plotted the charges along the length of the well. In the short-well regime, the borehole is approximately an equipotential surface and the charges are uniformly distributed; in the long well the charges decay with depth. What was surprising to us was the noticeable increase in charge accumulation that occurs near the bottom of the well. This is especially evident for the short well. Initially, we were suspicious and thought this might be due to problems with our numerical simulation; there was no obvious physical explanation that we were aware of. However, investigation into the literature revealed that the increase in charge density at the ends of a cylinder is a real physical effect, but an exact theoretical solution does still not appear to exist \citep{Griffiths1997} (see figure 4, in particular).

The results shown in Figure \ref{fig:kaufman_finite_well} have implications when testing approaches for reducing computational load by approximating a well with a solid cylinder or prism, as in \cite{Um2015}, or replacing the well with a distribution of charges, as in \cite{Weiss2016}. For a short well, the behaviour of the currents is independent of conductivity, so, as long as the borehole is approximated by a sufficiently conductive target, the behaviour of the fields and fluxes will be representative of the fine-scale model. However, as the length of the well increases, the cross-sectional conductance of the well becomes relevant as it controls the rate of decay of the currents in the well and thus the rate that currents leak into the formation. A similar result holds when a line of charges is used to approximate the well as a DC source; a uniform charge is suitable for a sufficiently short or sufficiently conductive well, whereas a distribution of charge which decays exponentially with depth needs to be considered for longer wells. Thus, when attempting to replace a fine-scale model of a well with a coarse-scale model, either with a conductivity structure or by some form of ``equivalent source'', validations should be performed on models that have the same length-scale as the experiment to ensure that both behaviors are being accurately modeled.

\subsection{Time Domain Electromagnetics}
\label{sec:TDEM}

\begin{figure}
    \begin{center}
    \includegraphics[width=0.95\columnwidth]{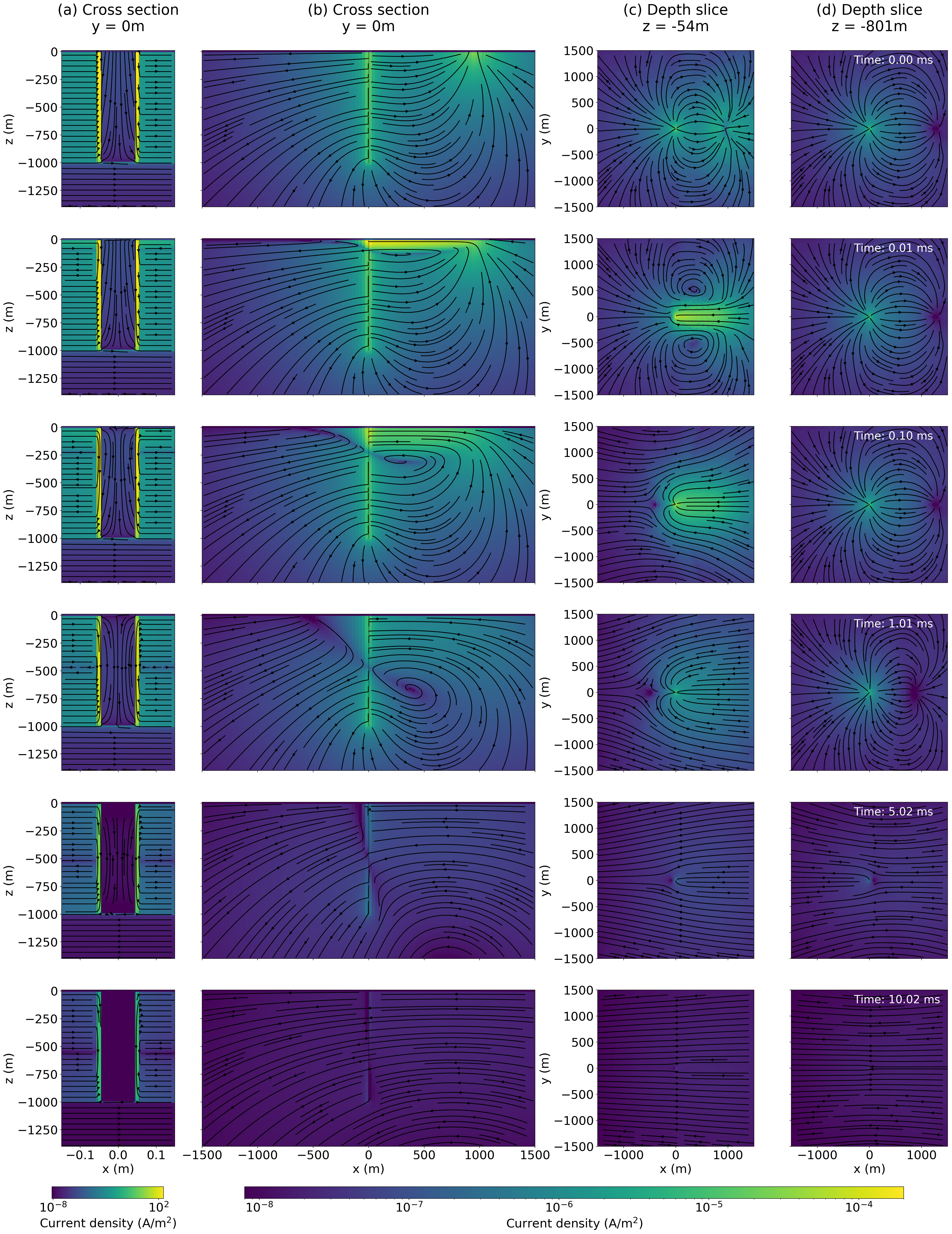}
    \end{center}
\caption{
    Current density for a time domain experiment where one electrode is connected to the top of the casing and a return electrode is on the surface, 1000m away.
    Six different times are shown, corresponding to each of the six rows; the times are indicated in the plots in panel (d).
    Panel (a) shows a zoomed-in cross section of the current density in the immediate vicinity of the steel cased well.
    Panel (b) shows a cross section through the half-space along the same line as the source-wire.
    Panels (c) and (d) show depth-slices of the currents at 54m and 801m depth.
}
\label{fig:TDEM_currents}
\end{figure}

In this example, we examine the behaviour of electric currents in an experiment where the casing is used as an ``extended electrode''. Although the initial investigations with casings centered around using a DC source, greater information about the subsurface can be had by employing a frequency or time domain source. A particular application is the monitoring of hydraulic fracturing proppant and fluids, or CO$_2$; this is active research carried out by many groups worldwide (e.g. \cite{Hoversten2015, Um2015, Puzyrev2017, Zhang2018} among others). The challenge is to have efficient and accurate forward modeling; solving the full Maxwell equations is much more demanding than solving the DC problem. For our simulation, a positive electrode is connected to the top of the casing and a return electrode is positioned 1 km away. The well has a conductivity of $10^6$ S/m and is 1 km long; it has an outer diameter of 10 cm and a 1 cm thick casing wall. The mud which infills the well has the same conductivity as the background, $10^{-2}$ S/m. The conductivity of the air is set to $10^{-5}$ S/m; in numerical experiments, we have observed that contrasts near or larger than $\sim 10^{12}$ S/m leads to erroneous numerical solutions. For this example, we will focus on electrical conductivity only and set the permeability of the well to $\mu_0$. A step-off waveform is used, and the currents within the formation are plotted through time in Figure \ref{fig:TDEM_currents}. Panel (a) shows a zoomed-in cross section of the casing, (b) shows a vertical cross section along the line of the wire (c) shows a horizontal depth slice at 50 m depth and (d) shows a depth slice at 800 m depth. The images in panels (b), (c) and (d) are on the same color scale.

We begin by examining Figure \ref{fig:TDEM_currents} (b), which shows the currents in the formation. At time $t=0s$, we have the DC solution. Currents flow away from the well, and eventually curve back to the return electrode. Immediately after shut-off, we see an image current develop in the formation. The image current flows in the same direction as the original current in the wire; this is opposite to currents in the formation, causing a circulation of current. The center of this circulation is visible as the null propagating downwards and to the right in Figure \ref{fig:TDEM_currents} (b). In Figure \ref{fig:TDEM_currents} (a), we see the background circulating currents being channeled into the well and propagating downwards. The depth range over which currents enter the casing depends upon time. At t=0.01 ms, the zero crossing, which distinguishes the depth between incoming and outgoing current in the casing, occurs at z=90 m, at t=0.1 ms it is at 225 m and by t=1 ms, the zero crossing approaches the midway point in the casing and is at 470 m depth. At later times, the downward propagation of this null slows as the image currents are channeled into the highly conductive casing; at 5 ms it is at 520 m depth, at 10 ms, 560 m depth and by 100 ms (not shown), it is at 800 m depth.

On the side of the well opposite to the wire, we also see a null develop; it is visible in the cross sections in panel (a). To help understand this, we examine the depth slices in panel (c). Behind the well, we see that as the image currents diffuse downwards and outwards, some of those currents are channeled back towards the well; this is visible in the depth slice at $10^{-4} s$. These channeled currents are opposite in direction to those the formation currents set up at t=0, which also are diffusing downwards and outwards; where these two processes intersect, there is a current shadow.

There are a number of points to highlight in this example. The first, which has been noted by several authors (e.g. \cite{Schenkel1994, Hoversten2015}), is that the casing helps increase sensitivity to targets at depth. This occurs by two mechanisms: (1) at DC, prior to shut-off, the casing acts as an ``extended electrode'' leaking current into the formation along its length; (2) after shut-off, it channels the image currents and increases the current density within the vicinity of the casing. The second point to note is that there are several survey design considerations raised by examining the currents: targets that are positioned where there is significant current will be most illuminated. If the target is near the surface and offset from the well, a survey where the source wire runs along the same line as the target will have the added benefits of the excitation due to the image currents. These benefits are twofold: (1) the passing image-current increases the current density for a period of time, and (2) the changing amplitude and direction of the currents with time generate different excitations of the target. this should provide enhanced information in an inversion, as compared to a single excitation that is available from a DC survey. For deeper targets in this experiment, the passing image current has diffused significantly, and thus it appears that the wire location has less impact on the magnitude of the current density with location. However, it is possible that increasing the wire-length could be beneficial. This extension is straightforward and could be examined with the provided script. There may also be added benefit by having the target positioned along the same line as the source wire, as at later times, the direction of current reverses, changing the excitation of the target. The final point to note from this example is that although this is a simple model, the behavior of the currents is not intuitive; visualizations of the currents, fields and fluxes, allow researchers to explore the basic physics and prompts new questions.

\subsection{Frequency Domain Electromagnetics Part 1: Comparison with scale model results}
\label{sec:FDEM_part1}

In the DC example, we discussed how charges are distributed along the well and currents flow into the formation. The time domain example extended the analysis of grounded sources, showed the potential importance of EM induction effects and illuminated the underlying physics. From a historical perspective, however, practical developments in EM were pursued in the frequency domain; the mathematics is more manageable in the frequency domain, and technological advances were being made in the development of induction well-logging tools \citep{Doll1949, Moran1962}. Although conductivity of the pipes is generally plays the most dominant role in attenuating the signal, the magnetic permeability is non-negligible \citep{Wait1977}; it is the product of the conductivity and permeability that appears in the description of EM attenuation. Also, the fact that permeable material becomes magnetized in the presence of an external field complicates the problem.
\cite{Augustin1989} is one of the first papers on induction logging in the presence of steel cased wells that aims to understand and isolate the EM response of the steel cased well. Using a combination of scale modeling and analytical mathematical modeling, they examined the impacts of conductivity and magnetic permeability on the magnetic field observed in the pipe. In this example and the one that follows, we attempt to unravel this interplay between conductivity and magnetic permeability.

The first experiment \cite{Augustin1989} discuss is a scale model using two different pipes, a conductive copper pipe and a conductive, permeable iron pipe; each pipe is 9 m in length. The copper pipe had an inner diameter of 0.063 m and a thickness of 0.002 m, while the iron pipe had a 0.063 m inner diameter and 0.0043 m wall thickness. A source-loop, with radius 0.6 m was co-axial with the pipe and in one experiment positioned at one end of the pipe (which they refer to as the ``semi-infinite pipe'' scenario). In another experiment the source loop is positioned at the midpoint of the pipe (which they refer to as the ``infinite pipe'' scenario); for both experiments, magnetic field data are measured as a function of frequency at the central axis of the pipe. Their results are presented in terms of a Field Strength Ratio (FSR), which is the ratio of the absolute value of the magnetic field at the receiver with the absolute value of the magnetic field if no pipe is present (Figure 3 in \cite{Augustin1989}). At low frequencies, for the data collected within the iron pipe, static shielding (FSR $<$ 1) was observed for the measurements where the receiver was in the plane of the source loop for both the ``infinite'' and ``semi-infinite'' scenarios. When the receiver was positioned within the pipe, 1.49 m offset from the plane of the source loop, static enhancement effects (FSR $>$ 1) were observed for both the infinite and semi-infinite scenarios. Using this experiment for context, we will compare the behaviour of our numerical simulation with the observations in \citep{Augustin1989} and examine the nature of the static shielding and enhancement effects.

\begin{figure}
    \begin{center}
    \includegraphics[width=0.6\columnwidth]{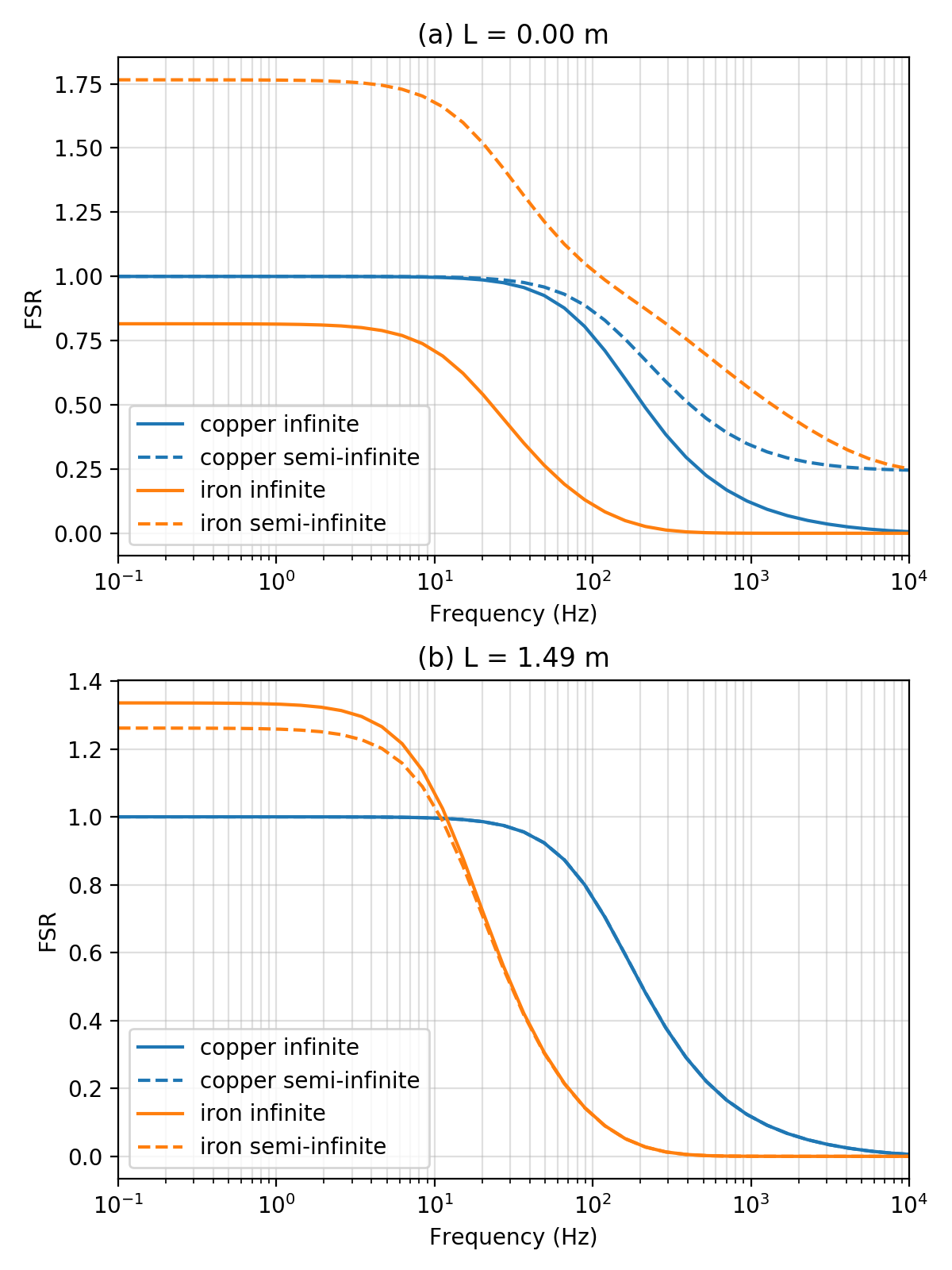}
    \end{center}
\caption{
    Field strength ratio (FSR), the ratio of the measured vertical magnetic field with the free space magnetic field, as a function of frequency for two different reciever locations.
    In (a), the reciever is in the same plane as the source, in (b), the reciever is 1.49m offset from the source.
}
\label{fig:AugustinFSR}
\end{figure}

For our numerical setup, the pipes are 9 m in length and have an inner diameter of 0.06 m. The copper pipe has a casing-wall thickness of 0.002 m and the iron pipe has a thickness of 0.004 m. Following the estimated physical property values from \cite{Augustin1989}, we use a conductivity of $3.5 \times 10^7$ S/m and a relative permeability of 1 for the copper pipe. For the iron pipe, a conductivity of $8.0 \times 10^6$ S/m and a relative permeability of 150 is used. A background conductivity of $10^4 ~\Omega$m is assumed. The computed FSR values for the axial magnetic field as a function of frequency are shown in Figure \ref{fig:AugustinFSR}.

Consider the response of the conductive pipe. At low frequencies, the FSR for the copper pipe (blue lines) is 1 for both the infinite (solid line) and semi-infinite (dashed line) scenarios, as the field inside the copper pipe is equivalent to the free-space field. With increasing frequency, eddy currents are induced in the pipe which generate a magnetic field that opposes the primary, causing a decrease in the observed FSR. When the source and receiver are in the same plane (L=0.00 m), the rate of decrease is more rapid in the infinite scenario than the semi-infinite. Since there is conductive material on both sides of the receiver in the infinite case, we would expect attenuation of the fields to occur more rapidly than in the semi-infinite case. This observation is consistent with Figure 3a in \cite{Augustin1989}. For the offset receiver (L=1.49 m), they observed a slight separation in the infinite and semi-infinite curves which we do not; however, they attributed this to potential errors in magnetometer position. Thus, overall, the numerical results for the copper pipe are in good agreement with the scale model results observed by \cite{Augustin1989}.

Next, we examine the response of the conductive, permeable pipe. In Figure \ref{fig:AugustinFSR}b, we observe a static enhancement effect (FSR $>$ 1) at low frequencies. The enhancement is larger in the infinite scenario than the semi-infinite scenario; this is in agreement with Figure 3b in \cite{Augustin1989}. There is however, a significant discrepancy between our numerical simulations and the scale model for the semi-infinite pipe when the source and receiver lie in the same plane(Figure \ref{fig:AugustinFSR}a). \cite{Augustin1989} observed a static shielding effect for both the infinite and semi-infinite scenarios, whereas we observe a static shielding for the infinite scenario, but a significant static enhancement for the semi-infinite case. To diagnose what might be the cause of this, we will examine the magnetic flux density in this region of the pipe.

\begin{figure}
    \begin{center}
    \includegraphics[width=\columnwidth]{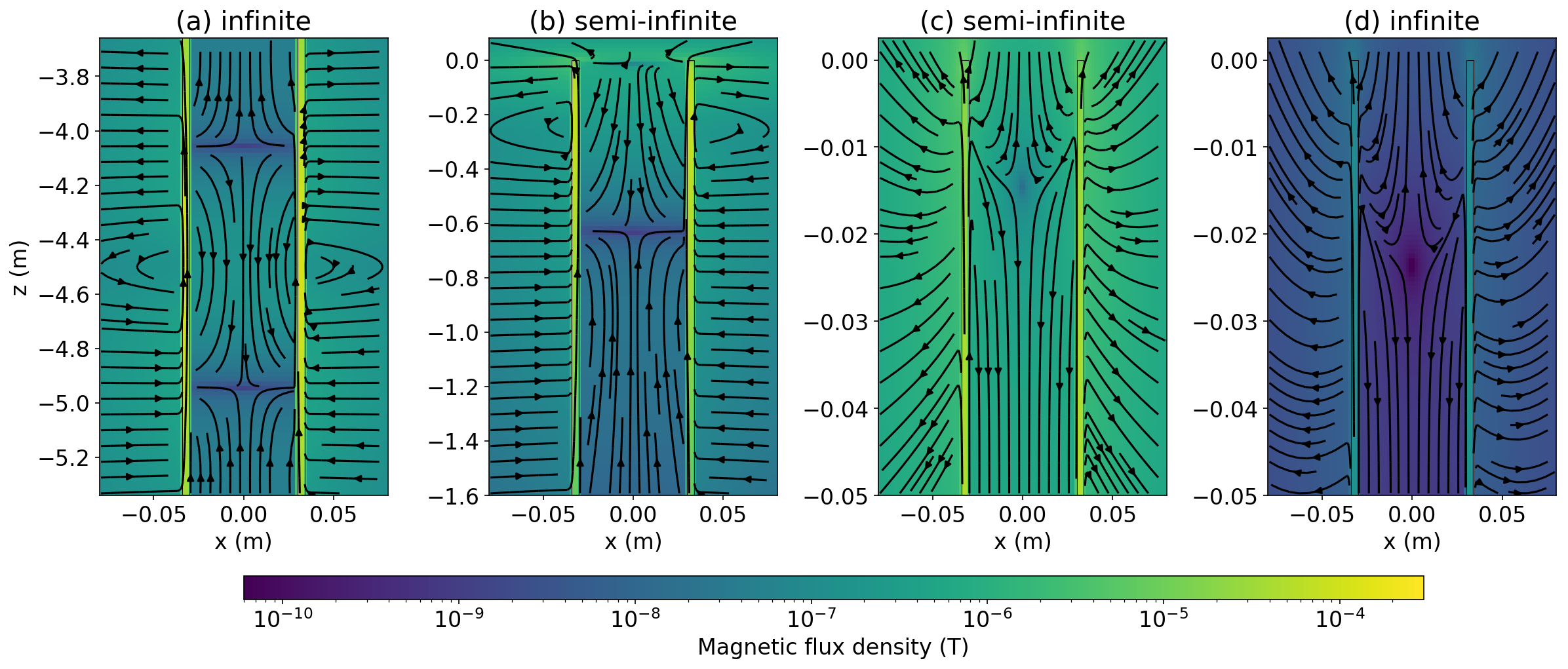}
    \end{center}
\caption{
    Magnetic flux density at 0.1Hz in the region of the pipe near the plane of the source for
    (a) the ``infinite'' pipe, where the source is located at -4.5m and the pipe extends from 0m to -9m,
    (b) a ``semi-infinite'' pipe, where the source is located at 0m and the pipe extends to -9m.
    In (c), we zoom in to the top 5cm of the ``semi-infinite'' pipe,
    and (d) shows the 5cm at the top-end of the ``infinite'' pipe.
}
\label{fig:AugustinBfields}
\end{figure}

In Figure \ref{fig:AugustinBfields}, we have plotted: (a) the secondary magnetic flux in the infinite-pipe scenario near the source (z=-4.5 m), (b) the secondary magnetic flux in the semi-infinite scenario (z=0 m for the source), and (c) top 5 cm of the semi-infinite pipe. All plots are at 0.1 Hz. The primary magnetic field is directed upwards within the regions we have plotted, so upward-going magnetic flux indicates a static enhancement effect, and downward-oriented magnetic flux indicates static shielding effects. In (a) we see a transition between the static shielding in the vicinity of the source to a static enhancement approximately 0.5 m above and below the plane of the source. Similarly in (b), we notice a sign-reversal in the z-component of the secondary magnetic flux at a depth of 0.6 m.

The behaviors observed in Figure \ref{fig:AugustinBfields} are quite comparable to Augustin et al.'s observation of a transition from shielding to enhancement occurring at distances greater than 0.8 m from the source. Numerical experiments show that the vertical extent of the region over which static shielding is occuring increases with increasing pipe diameter, and similarly increases with increasing loop radius while the magnitude of the effect decreases. This can be understood by considering how the pipe is magnetized; for a small loop radius, the magnetization is largely localized near the plane of the source and rapidly falls off with distance from the plane of the source. Localized, large amplitude magnetization causes the casing to act as a collection of dipoles around the circumference of the casing. As the radius of the loop increases, the magnetization spreads out along the length of the well resulting in longer, lower-amplitude dipoles, thus both increasing the extent of the region over which static shielding is occuring as well as decreasing its amplitude.

\begin{figure}
    \begin{center}
    \includegraphics[width=0.6\columnwidth]{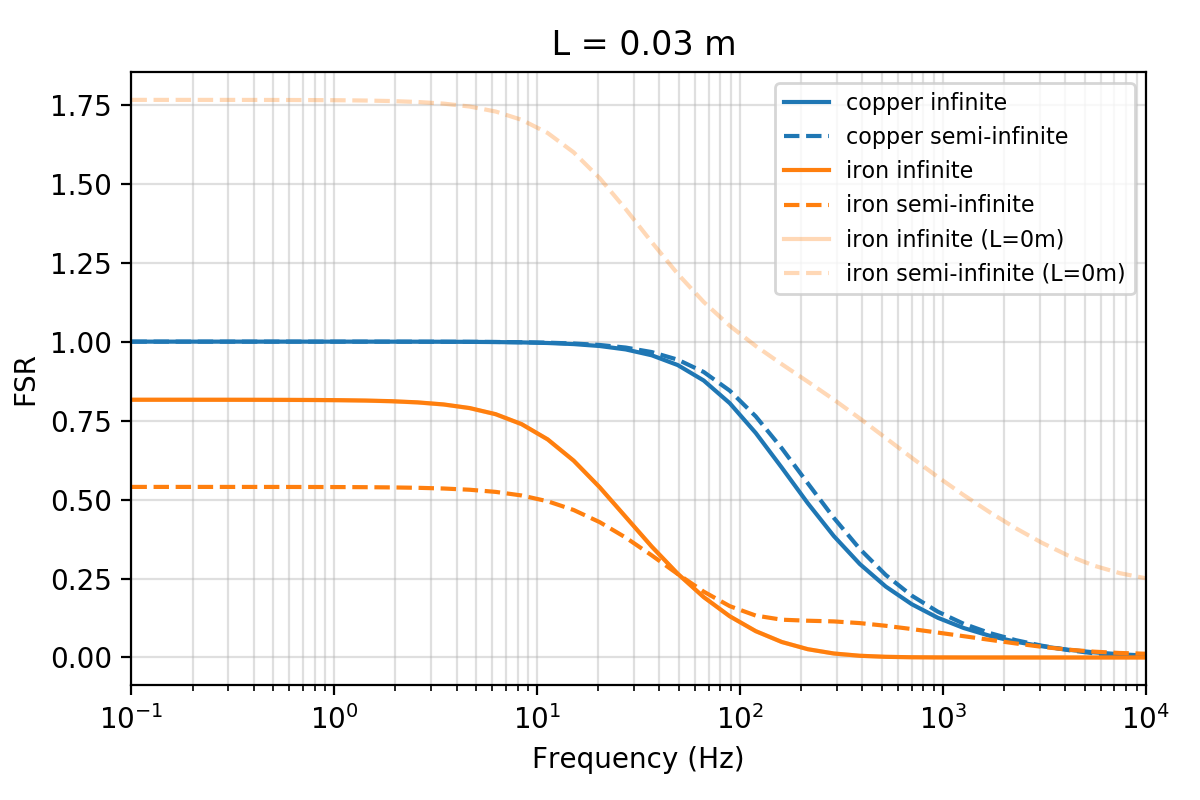}
    \end{center}
\caption{
    Field strength ratio, FSR, for a reciever positioned 3cm beneath
    the plane of the source. For comparison, we have plotted the
    FSR for the permeable pipe when the source and reciever lie in the same
    plane (L=0.00m) with the semi-transparent orange lines.
    Note that the infinite-pipe solutions for L=0.03m and L=0.00m overlap.
}
\label{fig:Augustin3cm}
\end{figure}

This explains the nature of the static enhancement and static shielding effects, but to explain the discrepancy between the static shielding observed in the semi-infinite pipe when L=0 m by Augustin et al., and the static enhancement we observe in Figure \ref{fig:AugustinFSR}a, we examine the magnetic flux density in the top few centimeters of the pipe. Figure \ref{fig:AugustinBfields}c shows the top 5 cm of the secondary magnetic flux in the semi-infinite pipe; the source is in the z=0 m plane.  Zooming in reveals there is yet another sign reversal near the end of the pipe. This is evident even in the infinite-pipe scenario (Figure \ref{fig:AugustinFSR}d), where the source is offset by several meters from the end of the pipe. This edge-effect perhaps bears some similarities to what we observed in Figure \ref{fig:kaufman_finite_well}b, where we saw a build up of charge near the end of the pipe in the DC scenario. At the end of the pipe, we encounter the situation where the normal component of the flux ($\vec{j}, \vec{b}$) from the pipe to the background needs to be continuous both in the radial and vertical directions at the end of the pipe as does the tangential component of the fields ($\vec{e}, \vec{h}$). The interplay of these two constraints at the end of the pipe results in more complexity in the resultant fields and fluxes. Within the span of a few centimeters we transition from static enhancement at the top of the pipe to a static shielding further down. An error as small as a few centimeters in the position of the magnetometer causes a reversal in behavior; in Figure \ref{fig:Augustin3cm}, we have plotted the FSR for a magnetometer positioned 3cm beneath the plane of the source, and the static-shielding behavior observed for the semi-infinite pipe is much more aligned with that observed in Figure 3a in \cite{Augustin1989}.

\subsection{Frequency Domain Electromagnetics Part 2: Conductivity and permeability in the inductive response of a well}
\label{sec:FDEM_part2}

The experiments shown in the previous section revealed some insights into the complexity of the fields within the pipe and illustrated the role of permeability in the character of the responses at low frequency. Next, we move to larger scales and examine the role of conductivity and permeability in the responses we observe in the borehole.

In this example, we consider a 2 km long well with an outer diameter of 10 cm and thickness of 1 cm in a whole-space which has a resistivity of $10^4$ $\Omega$m. A loop with radius 100 m is coaxial with the well and positioned at the top-end of the well. A receiver measuring the z-component of the magnetic flux density is positioned 500 m below the transmitter loop, along the axis of the well. We will consider both time domain and frequency domain responses.

In electromagnetics, it is often the product of permeability and conductivity that we consider to be the main controlling factor on the EM responses. To assess the contribution of each to the measured responses, we will investigate two scenarios. In the first, the well has a conductivity of $10^8$ S/m and a relative permeability of 1, and in the second, the well has a conductivity of $10^6$ S/m and a relative permeability of 100; thus the product of conductivity and permeability is equivalent for both wells.

\begin{figure}[htb]
    \begin{center}
    \includegraphics[width=0.6\columnwidth]{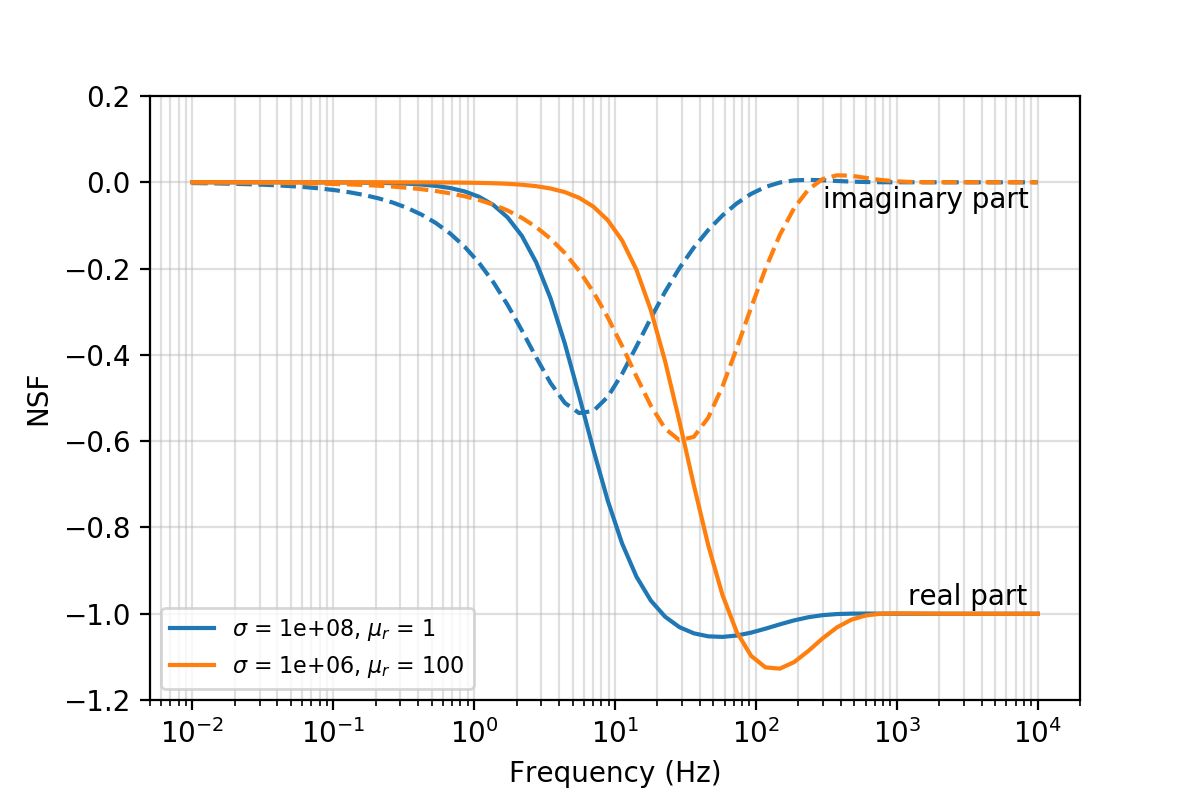}
    \end{center}
\caption{
    Normalized secondary field, NSF, as a function of frequency for two wells.
    The NSF is the ratio of the secondary vertical magnetic field with the primary magnetic field at the reciever location (z=-500m);
    the primary is defined as the whole-space primary.
}
\label{fig:fdemNSF}
\end{figure}

Similar to the analysis done by \cite{Augustin1989} when looking at the role of borehole radius in the behaviour of the magnetic response (e.g. figure 8), we will examine the normalized secondary field (NSF) which is the ratio of the secondary field with the amplitude of the primary, where the primary is defined to be the free-space response. In Figure \ref{fig:fdemNSF}, we have plotted the normalized secondary field for the two pipes considered, the conductive pipe (blue) and the conductive, permeable pipe (orange). Let's start by examining the conductivity response in Figure \ref{fig:fdemNSF}. Where the value of the NSF is zero, the primary dominates the response; this is the case at low frequencies where induction is not yet contributing to the response. As frequency increases, currents are induced in the pipe which generate a secondary magnetic field that opposes the primary, hence the NSF becomes negative. When the real part of the NSF (solid line) is -1, the secondary magnetic field is equal in magnitude but opposite in direction to the free-space primary and the measured real field is zero. Values less than -1 indicate a sign reversal in the real magnetic field. Similarly, when the imaginary part of the response function goes above zero, there is a sign reversal in the imaginary component. Note that these sign reversals occur even in a half-space and are a result of sampling the fields within a conductive medium; in this case the receiver was 500 m below the surface.

\begin{figure}
    \begin{center}
    \includegraphics[width=\columnwidth]{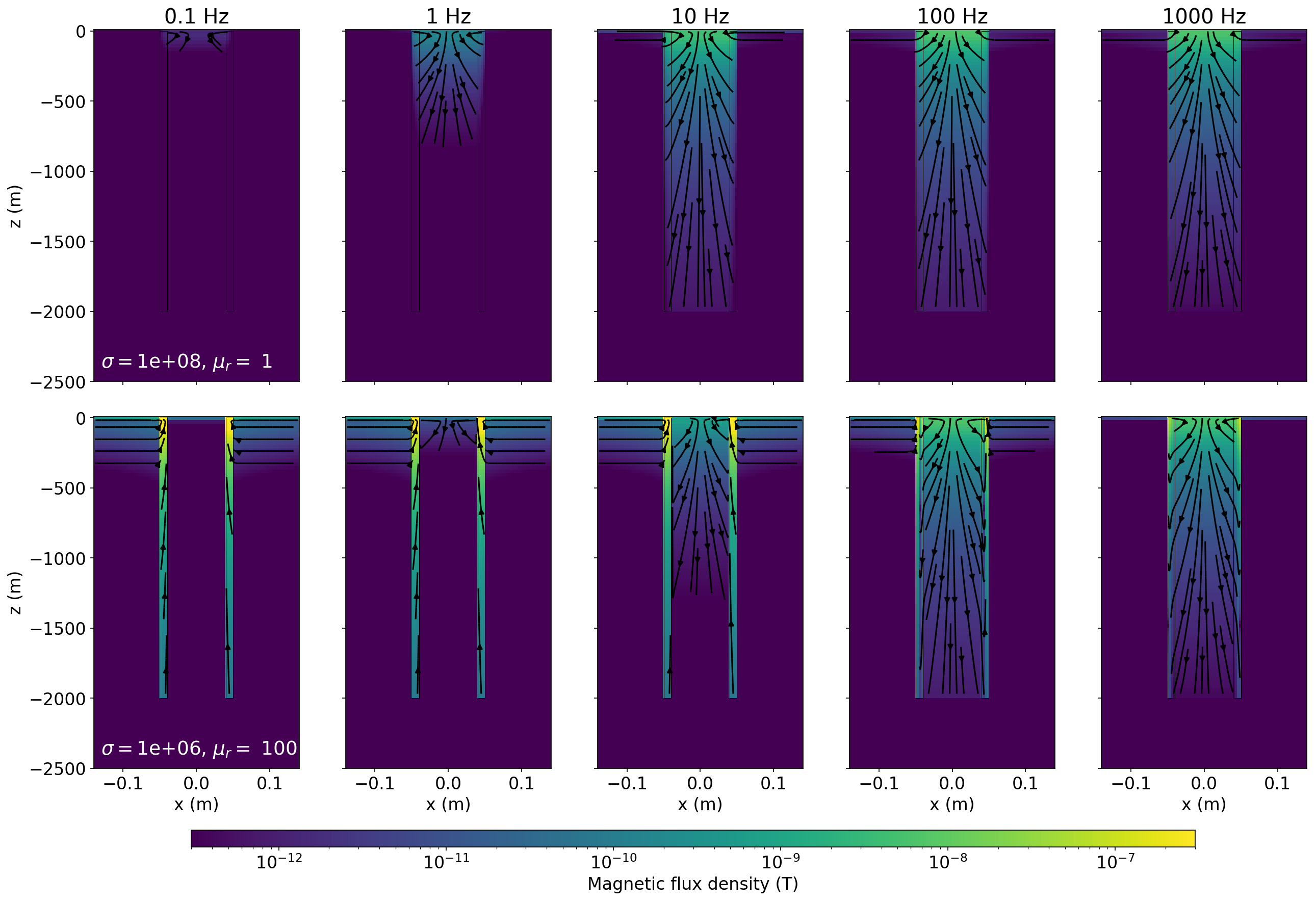}
    \end{center}
\caption{
    Secondary magnetic flux density (with respect to a whole-space primary) at five different frequencies for a conductive pipe (top row)
    and for a conductive, permeable pipe (bottom row).
}
\label{fig:bfdem}
\end{figure}

As compared to the conductive pipe, the frequency at which induction sets in is higher for the conductive, permeable pipe. We also notice that the amplitude variation of both the imaginary and real parts is larger for the permeable pipe. To examine the contribution of conductivity and permeability to the responses, we have plotted the real part of the secondary magnetic flux density, $\mathbf{b}$, in Figure \ref{fig:bfdem}. The top row shows the response within the conductive pipe and the bottom row shows the conductive, permeable pipe. The primary magnetic flux is oriented upwards and we can see that all of the secondary fields generated are oriented downwards. Similar to the previous example, we see that at low frequencies, there is magnetostatic response due to the permeable pipe. However, due to the larger length scales of the source loop and the casing in this example, there is no measurable contribution at the receiver. At 1 Hz, we can see that induction is starting to contribute to the signal for the conductive pipe, while for the permeable pipe, it is not until $\sim$10 Hz that we begin to observe the contribution of induction. At 100 Hz, the secondary magnetic field is stronger in amplitude than the primary, and the NFS is less than -1 for both the conductive and permeable pipes. The amplitude of the secondary within the permeable pipe is stronger than that in the conductive pipe. At 1000 Hz, we have reached the asymptote of NSF=-1 for both the conductive and permeable pipes; the secondary magnetic flux is equal in magnitude but opposite in direction to the primary.


\begin{figure}[htb]
    \begin{center}
    \includegraphics[width=0.6\columnwidth]{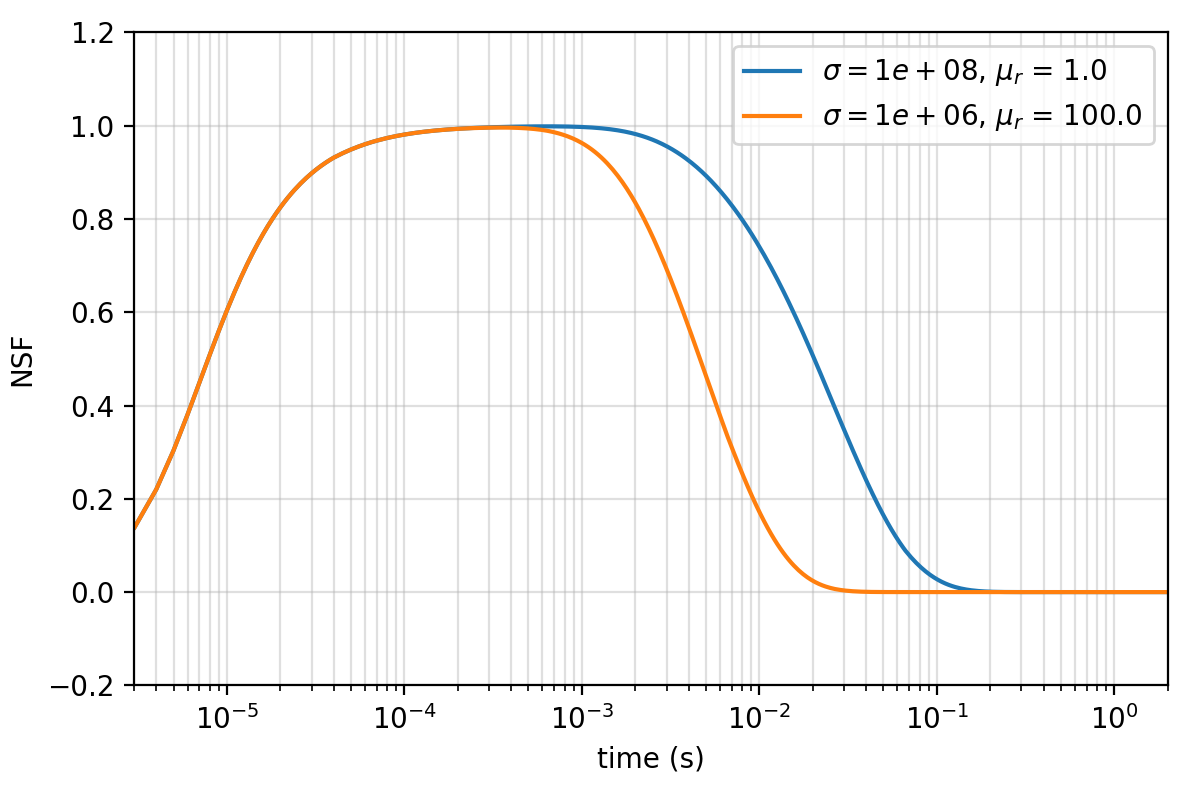}
    \end{center}
\caption{
    Normalized secondary field (NSF) through time.
    In the time-domain, we compute the NSF by taking the difference between the total magnetic flux at the reciever and the whole-space response
    and then taking the ratio with the whole-space magnetic flux prior to shutting off the transmitter.
}
\label{fig:tdemNSF}
\end{figure}

\begin{figure}
    \begin{center}
    \includegraphics[width=\columnwidth]{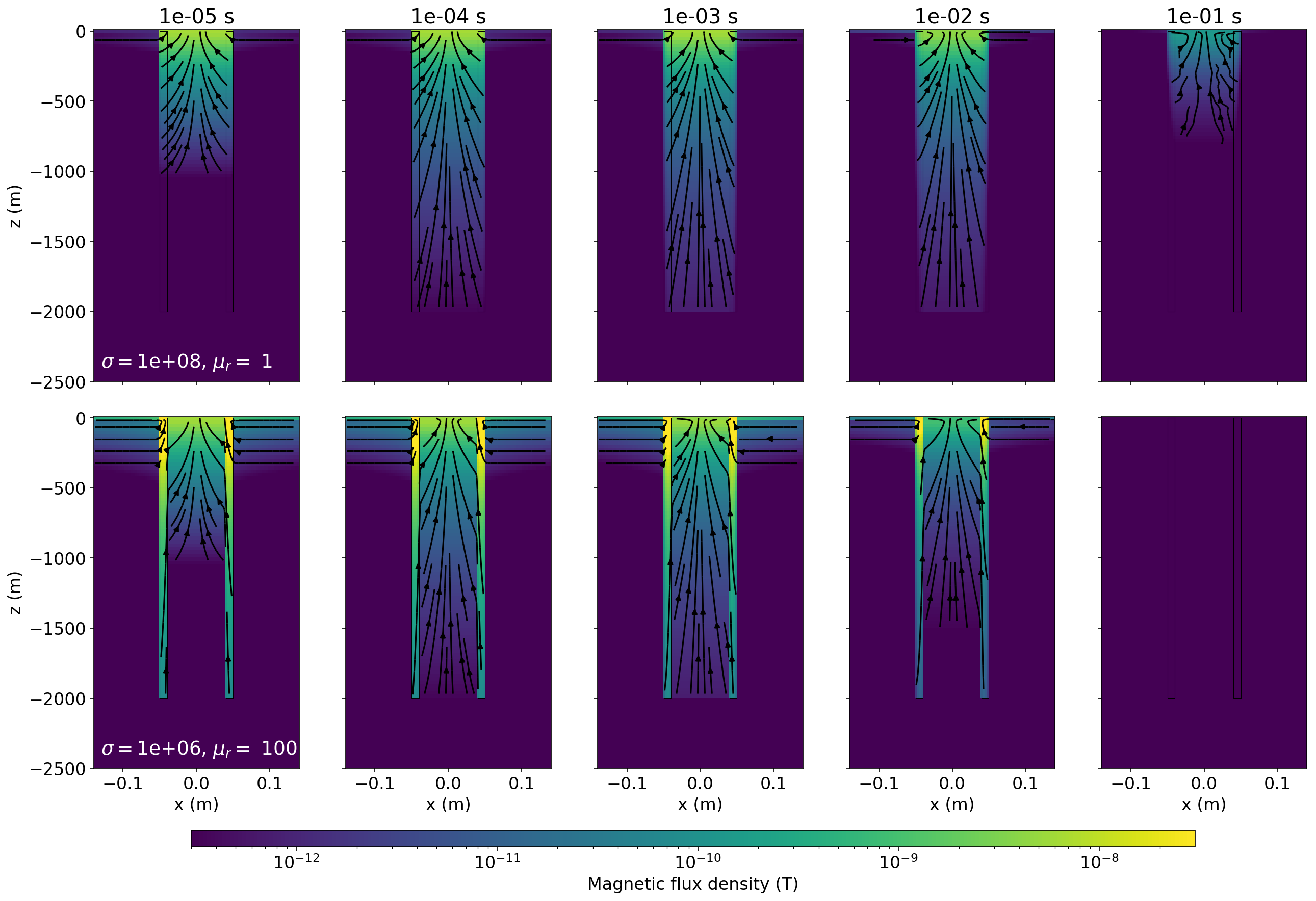}
    \end{center}
\caption{
    Secondary magnetic flux density for a conductive well (top row) and a conductive, permeable well (bottom row) through time.
    The source waveform is a step-off waveform.
}
\label{fig:btdem}
\end{figure}

Conducting a similar experiment in the time domain, we can compare the responses as a function of time. For this experiment, a step-off waveform is employed and data are measured after shut-off, the NSF is plotted in Figure \ref{fig:tdemNSF}. Note here that the secondary field is in the same direction as the primary, so after the source has been shut off, the secondary field is oriented upwards, as shown in Figure \ref{fig:btdem}. Shortly after shut-off, the rate of increase in the secondary field is the same for both the conductive and the conductive, permeable wells. A maximum normalized field strength of approximately 1 is reached for both cases. The responses begin to differ at $10^{-3}$ s where the conductive well maintains a NFS $\sim 1$ for approximately 1 ms longer than the permeable well before the fields decay away.


It is important to note that although the product of the conductivity and permeability is identical for these wells, the geometry of the well and inducing fields results in different couplings for each of the parameters. For a vertical magnetic dipole source, the electric fields are purely rotational while the magnetic fields are primarily vertical. An approximation we can use to understand the implications of these geometric difference is to assume the inducing fields are uniform (e.g. the radius of the source loop is infinite) and to examine the conductance and permeance of the pipe. For rotational electric fields, the conductance is
\begin{equation}
    \mathcal{S} = \sigma \frac{t L}{2 \pi r}
    \label{eq:conductance}
\end{equation}
where $t$ is the thickness of the casing, $r$ is the radius of the casing and $L$ is the length-scale of the pipe segment contributing to the signal. For vertical magnetic fields, the permeance is
\begin{equation}
    \mathcal{P} = \mu \frac{ t 2 \pi r}{L}
    \label{eq:permeance}
\end{equation}
As the length-scale, $L$, is larger than the circumference of the pipe ($2\pi r$) the geometric contribution to the conductance is larger than that to the permeance.

An important take-away from this example is that the contributions of conductivity and permeability to the observed EM signals are not simply governed by their product. The geometry of the source fields plays an important role in how each contributes. Thus to accurately model conductive, permeable pipes, over a range of frequencies or times, a numerical code must allow both variable conductivity and variable permeability to be considered.

\section{Discussion}
The behavior of EM fields and fluxes in the presence of highly conductive, permeable, steel-cased wells is often non-intuitive. In DC resistivity experiments, we demonstrated that there is a charge build-up near the end of the well. This has not previously been discussed in the geophysics literature, but it was recognized by \cite{Griffiths1997}. In practice, such a charge buildup might be consequential in an inversion as it will alter the sensitivities; this is an avenue for future research.

Moving to EM experiments complicates matters in two ways: (1) the fields and fluxes vary through time introducing inductive phenomena, and (2) variable magnetic permeability alters the fields and fluxes. With respect to magnetic permeability, \cite{Augustin1989} noticed ``static shielding'' and ``static enhancement'' effects in a scale-model experiment with an iron pipe subject to an inductive source. They observed that ``the truncated pipe is more effective at shielding static and low-frequency fields than the infinite pipe,'' however, they offered no explanation as to why this is the case. In the numerical experiment in section \ref{sec:FDEM_part1}, we were able to replicate the nature of their data, and furthermore, we demonstrated that near the ends of the pipe, the magnetic fields change rapidly over short length-scales. Although this may seem to be a detail that is unimportant unless measurements are being made near the end of a pipe, it demonstrates some of the complexity that is introduced when both conductivity and permeability are significant in a model. To date, the geophysics literature that considers the magnetic permeability of well-casings does so primarily in the context of inductive sources; very little research examines the impact of permeability on grounded-source experiments. Thus, there are open questions about how magnetic permeability alters the currents and impacts the data in such experiments. Building an understanding of these impacts is critical both for assessing feasibility of using EM methods to delineate targets of interest as well as for developing strategies to reduce the computational cost of 3D simulations which include steel-cased wells.

In many settings where DC or EM experiments are being considered, wells are deviated or horizontal, and several wells may be present. The cylindrical discretization strategy presented here does not accommodate such geometries. Recent advancements such as the hierarchical finite element approach presented in \cite{Weiss2017} make modeling these complex scenarios feasible for DC resistivity. However, it is not clear that approaches that work at DC will be suitable for EM. For example, \cite{Weiss2017} points out several challenges that arise when considering the inherently more complex PDE governing EM, and in particular, points out that it is unclear how to include magnetic permeability in a hierarchical approach. For EM, other avenues for performing simulations include the upscaling approach suggested by \cite{Schwarzbach2018, Caudillo-Mata2017} in which one inverts for the conductivity and permeability of a coarse-scale model of the casing, as well as the method of moments approach taken by \cite{Kohnke2017}. Irrespective of the strategy that is taken, it is important to have numerical tools that yield accurate, computationally efficient simulations and that are easy to use. Although tools such as COMSOL are versatile, the cognitive overhead for a researcher to set up a simple simulation to test their understanding of the physics is significant. The software presented here aims to bridge that gap and serve as a resource for researchers to calibrate their understanding of the physics, as well as to assess the assumptions that new approaches are making and to benchmark their accuracy.
\section{Summary and Outlook}

We developed software for solving Maxwell's equations on 2D and 3D cylindrical meshes. Variable electrical conductivity and magnetic permeability are considered. The 2D solution is especially computationally efficient and has a large number of practical applications. When cylindrical symmetry is not valid, the 3D solution can be implemented; a judicious design of the mesh can often generate a problem with fewer cells than would be required with a tensor or OcTree mesh, thus reducing the computational cost of a simulation. We demonstrated the versatility of the codes by modeling the electromagnetic fields that result when a highly conductive and permeable casing is embedded in the earth.

We presented a number of different experiments involving DC, frequency-domain, and time-domain sources. The first two examples considered a simple DC resistivity experiment. In the first, we demonstrated that the numerically obtained currents, electric fields, and charges emulated those predicted by the asymptotic analysis in \cite{Kaufman1990} for long wells. The second example looked at the transition in behavior of currents and charges between short and long wells. Even in this relatively simple example, the physics was more complex than we originally anticipated.

In the subsequent examples, we considered electromagnetic experiments. The third example presented a grounded-source time-domain experiment and showed the distribution of currents in the formation through time. It showed that the steel-casing can help excite a target at depth by two mechanisms: (1) the casing provides a high-conductivity pathway for bringing DC currents to depth, and (2) the casing channels the image current that is created after shut-off of the source. With respect to survey design, one consequence of the second point is that there may be an advantage to positioning the wire and return electrode along the same line as where the target is expected to be located. In this way, the current direction is reversed as the image current passes; the target is thus excited from multiple excitation directions and the resultant data can be beneficial in an inversion.

The final two examples incorporated magnetic permeability in the simulations. We showed that for a conductive and permeable casing, excited by a circular current source, there is a complicated magnetic field that occurs in the top few centimeters of the pipe. Furthermore, the role of conductivity and permeability in the observed responses is more complex than their product; the source geometry and coupling with the casing are important to consider. In each of the examples, the ability to plot the charges, fields, and fluxes was of critical importance; these ground our understanding of the physics and provide a foundation for designing a field survey.

The software implementation is included as a part of the \texttt{SimPEG} ecosystem. \texttt{SimPEG} also includes finite volume simulations on 3D tensor and OcTree meshes as well as machinery for solving inverse problems. Thus, the cylindrical codes can be readily connected to an inversion and additionally, simulations and inversions of more complex 3D geologic settings can be achieved by coupling the cylindrical simulation with a 3D tensor or OcTree mesh using a primary-secondary approach (e.g. example 3 in \cite{Heagy2017}). Beyond modeling steel cased wells, we envision that the 3D cylindrical mesh could prove to be useful in conducting 3D airborne EM inversions where a domain-decomposition approach, similar to that described in \cite{Yang2014}, is adopted.

\texttt{SimPEG} and all of the further developments described in this paper are open source and freely available. The examples have been provided as Jupyter notebooks. This not only allows all of the figures in the paper to be reproduced, but provides an avenue by which the reader can ask questions, change parameters, and use resultant images to confirm (or not) their presumed outcome. We hope that our efforts to make the software and examples accessible promotes the utility of this work for the wider community.

\section{Acknowledgements}
The authors thank Michael Commer and Christoph Schwarzbach for providing the simulation results shown in Figure \ref{fig:commer_results} and for permission to distribute them.
We also thank Thibaut Astic and Dikun Yang for their suggestions and input on the early draft of this paper. Finally, we are grateful to Rowan Cockett, Seogi Kang and the rest of the \texttt{SimPEG} community for their discussions and efforts on the development of the \texttt{SimPEG}.

We are also grateful to Raphael Rochlitz and the two other anonymous reviewers for their critiques and suggestions which improved the quality of the manuscript.

The funding for this work is provided through the Vanier Canada Graduate Scholarships Program.

\section{Computer Code Availability}
All of the software used in this paper is open source and was made available in 2018. The examples are provided as Jupyter notebooks and the code is written in Python. The code and installation instructions are available at https://github.com/simpeg-research/heagy-2018-emcyl and have been archived at https://doi.org/10.5281/zenodo.1220427. The main developer is Lindsey Heagy (email: lheagy@eos.ubc.ca, phone: (604) 836-2715).


\section*{References}
\bibliographystyle{elsarticle-harv}
\bibliography{refs.bib}\end{document}